\documentclass[preprint,sort&compress,12pt]{elsarticle}

\usepackage{amssymb}
\usepackage{amsthm}
\usepackage{amsmath}
\usepackage{mathtools}
\usepackage{mathrsfs}
\usepackage{algorithm}
\usepackage[algo2e]{algorithm2e}
\usepackage{algpseudocode}
\usepackage{array}
\usepackage{multirow}
\usepackage{listings}
\usepackage{tabu}
\usepackage{booktabs}
\usepackage{enumerate}
\usepackage{fullpage}
\usepackage{float}
\usepackage{xcolor}
\usepackage[colorinlistoftodos]{todonotes}
\usepackage{bbm}
\usepackage{bm}
\usepackage[colorlinks=true]{hyperref}
\usepackage{url}
\usepackage{textcomp}
\usepackage{gensymb}
\usepackage{soul}
\usepackage{lineno}
\usepackage{graphicx}
\usepackage{subfigure}
\usepackage{float}
\usepackage{caption}
\usepackage{tikz}
\usetikzlibrary{shapes}

\theoremstyle{definition}

\theoremstyle{remark}

\biboptions{numbers,comma,round,square}
\graphicspath{ {./figs/} }

\journal{Elsevier}

\begin{document}
\begin{frontmatter}

\title{Neural Differentiable Modeling with Diffusion-Based Super-resolution for Two-Dimensional Spatiotemporal Turbulence}



\author[ndAME,ndECI,ndEnergy]{Xiantao Fan}
\author[ndAME]{Deepak Akhare}
\author[ndAME,ndECI,ndEnergy,ndLucy]{Jian-Xun Wang\corref{corxh}}

\address[ndAME]{Department of Aerospace and Mechanical Engineering, University of Notre Dame, Notre Dame, IN}
\address[ndECI]{Environmental Change Initiative, University of Notre Dame, Notre Dame, IN}
\address[ndEnergy]{Center for Sustainable Energy (ND Energy), University of Notre Dame, Notre Dame, IN}
\address[ndLucy]{Lucy Family Institute for Data \& Society, University of Notre Dame, Notre Dame, IN}
\cortext[corxh]{Corresponding author. Tel: +1 540 3156512}
\ead{jwang33@nd.edu}

\begin{abstract}

Simulating spatiotemporal turbulence with high fidelity remains a cornerstone challenge in computational fluid dynamics (CFD) due to its intricate multiscale nature and prohibitive computational demands. Traditional approaches typically employ closure models, which attempt to represent small-scale features in an unresolved manner. However, these methods often sacrifice accuracy and lose high-frequency/wavenumber information, especially in scenarios involving complex flow physics. In this paper, we introduce an innovative neural differentiable modeling framework designed to enhance the predictability and efficiency of spatiotemporal turbulence simulations. Our approach features differentiable hybrid modeling techniques that seamlessly integrate deep neural networks with numerical PDE solvers within a differentiable programming framework, synergizing deep learning with physics-based CFD modeling. Specifically, a hybrid differentiable neural solver is constructed on a coarser grid to capture large-scale turbulent phenomena, followed by the application of a Bayesian conditional diffusion model that generates small-scale turbulence conditioned on large-scale flow predictions. Two innovative hybrid architecture designs are studied, and their performance is evaluated through comparative analysis against conventional large eddy simulation techniques with physics-based subgrid-scale closures and purely data-driven neural solvers. The findings underscore the potential of the neural differentiable modeling framework to significantly enhance the accuracy and computational efficiency of turbulence simulations. This study not only demonstrates the efficacy of merging deep learning with physics-based numerical solvers but also sets a new precedent for advanced CFD modeling techniques, highlighting the transformative impact of differentiable programming in scientific computing.


\end{abstract}

\end{frontmatter}

\section{Introduction}
\label{sec:intro}

The simulation of turbulent fluid flows is crucial for advancing our understanding and enhancing predictive capabilities across a wide range of applications, from aerospace engineering to environmental sciences. Despite significant advancements in scientific computing and numerical modeling, achieving an effective and efficient predictive simulation of turbulence remains a substantial challenge, primarily arising from the inherent multi-scale nature of turbulence and the significant computational demands required for accurately capturing these scales. The direct numerical simulation (DNS), which resolves all scales of turbulence structures, is only feasible for very limited simple flow configurations. For more complex scenarios, especially those involving high Reynolds numbers, the computational costs for DNS exceed practical limits, prompting the need for alternative approaches.  Methods such as Reynolds-Averaged Navier-Stokes (RANS) models and Large Eddy Simulations (LES) are viable solutions by approximating the effects of unresolved turbulence through closure modeling, effectively balancing computational demands with simulation accuracy~\cite{durbin2018some}. Nonetheless, these strategies, especially LES, continue to present substantial computational challenges for practical engineering applications. Moreover, the generalizability of closure models is often constrained to limited flow conditions, leading to large prediction errors in many practical engineering flows that exhibit features such as flow separations, strong pressure gradients, and pronounced mean-flow curvatures.

The exploration of alternatives to conventional CFD methods has been significantly advanced by the emergence of scientific machine learning (SciML), fueled by the increasing availability of high-quality data. This has led to a surge in SciML applications to enhance or potentially replace traditional CFD solvers, offering the promise of accelerated computations and improved predictive performance. The development of deep learning (DL) based surrogate or reduced-order models for fluid flows represents a notable progress. These models leverage advanced DL architectures such as autoencoding-based convolutional neural networks (CNN)~\cite{morimoto2021convolutional,guastoni2021convolutional}, graph neural networks (GNN)~\cite{pfaff2020learning,han2022predicting}, Fourier neural operators (FNO)~\cite{li2020fourier,li2023long}, DeepONet~\cite{lu2021learning,demo2023deeponet}, among others, showing initial success in simulating canonical fluid dynamics problems. Despite the promise, the application of purely data-driven DL surrogate models to complex turbulent flows presents substantial challenges due to the inherent chaos and complexity of such scenarios. These models often struggle with maintaining long-term forecasting stability and reliability, as the chaotic dynamics of turbulence lead to rapid accumulation of prediction errors. Additionally, their ``black-box'' nature demands extensive labeled datasets for training, a prerequisite that is rarely met in most practical situations. Consequently, these purely data-driven DL models face difficulties in maintaining accuracy and generalizing beyond the training regimes.

Hybrid learning approaches, which synergize machine learning (ML) with physics-based modeling, emerge as a strategic solution to the pitfalls of purely data-driven black-box ML methods. This strategy aims to leverage the learning capabilities of ML and theoretical foundation of fluid mechanics, potentially reducing the reliance on extensive training datasets and enhancing model generalizability. One approach in this direction involves utilizing the governing physics as penalty terms to constrain the training process of the ML models, e.g., physics-informed neural networks (PINNs)~\cite{karniadakis2021physics,cai2021physics}, where the governing partial differential equations (PDEs) can be constructed by automatic differentiation (AD)~\cite{raissi2019physics,sun2020surrogate,sun2022bayesian,sun2020physics} or numerical approximation~\cite{gao2021phygeonet,gao2022physics,ren2021phycrnet}. This approach has gained popularity for its ease of implementation and effectiveness in solving many canonical PDE problems~\cite{ren2024seismicnet,arzani2021uncovering,movahhedi2023predicting,li2022physics,kharazmi2021inferring}. However, their efficacy in navigating more complex fluid dynamics scenarios, such as those involving irregular three-dimensional geometries, turbulent and multi-scale phenomena, remains challenging, primarily due to the complexities of balancing multiple objectives during the training process~\cite{krishnapriyan2021characterizing,wang2022and}. An alternative approach of hybrid modeling is to merge ML with an established physics-based numerical solver. This integration aims to directly enhance the conventional solver's capability to handle complex fluid dynamics by introducing ML-derived insights into the simulation process. By doing so, it not only preserves the fundamental physics represented by traditional numerical solvers but also leverages the predictive power of ML to assimilate available data and improve performance. Specifically in turbulence modeling, ML models have been embedded into traditional CFD solvers for capturing unresolved turbulence through data-driven closure~\cite{ling2016reynolds,weatheritt2017hybrid,wang2017physics,wang2019prediction,maulik2019subgrid} or wall modeling~\cite{yang2019predictive, zhou2021wall, lozano2020self}. The majority of the existing literature focuses on learning the constitutive relationship between unresolved subgrid-scale (SGS) terms and mean-flow features in an \textit{a priori} manner using various ML models, such as symbolic regression~\cite{weatheritt2017hybrid}, random forest~\cite{wang2017physics}, and deep neural networks (DNN)~\cite{ling2016reynolds}. Despite achieving some success, particularly in encoding physical constraints (e.g., Galilean invariance, symmetry, divergence-free) through innovative feature engineering and ML architecture design, these models often fall short in ensuring \textit{a posteriori} accuracy and generalization across different flow conditions. Challenges include the stability of integrating offline-trained closure models with established RANS/LES frameworks~\cite{wu2019reynolds,guan2022stable} and the need for direct, accurately computed training labels like Reynolds stress or SGS stress, which are not always readily available or cannot be easily obtained~\cite{mcconkey2021curated,zhang2022ensemble}.

Addressing the challenges mentioned earlier, the trend toward a unified hybrid modeling via differentiable programming is gaining recognition~\cite{baydin2018automatic}. This innovative strategy enables a holistic optimization/training of both DL and numerical modeling components within a unified learning framework, which allows for direct interaction and feedback between these components, enhancing the overall model's ability to predict complex phenomena with greater accuracy and reliability. Through the integration of differentiable physics-based solvers and advanced DNNs, significant advances have been made across various scientific fields, demonstrating its effectiveness and robustness in learning physics with sparse and indirect training data~~\cite{mensch2018differentiable,innes2019differentiable,belbute2020combining, kochkov2021machine,list2022learned,akhare2023physics,akhare2023diffhybrid,akhare2023probabilistic,fan2024differentiable,liu2024multi}. For example, a JAX-based fully differentiable CFD solver has been integrated with CNNs for more effective and efficient coarse-grained simulations of 2D flows~\cite{kochkov2021machine}. Adjoint methods have been used to enable end-to-end learning of turbulence closure in RANS and LES simulations, aiming to achieve improved \emph{a posterior} prediction performance~\cite{macart2021embedded,strofer2021end,shankar2023differentiable,shankar2023differentiableturbulence}. Wang and co-workers have introduced the hybrid learning framework from a different perspective, known as \emph{neural differentiable modeling}, which is based on the profound relationship between the numerical representation of PDE operators and neural network architectural components, such as convolution layers, graph kernels, and residual connections~\cite{liu2024multi}. From this perspective, the traditional numerical solvers can be viewed as specialized neural networks predefined by established physics and their numerical representations. The effectiveness of the neural differentiable modeling framework has been demonstrated in learning spatiotemporal dynamics of fluids~\cite{liu2024multi}, fluid-structure-interactions~\cite{fan2024differentiable}, and manufacturing processes of composite materials~\cite{akhare2023physics,akhare2023diffhybrid,akhare2023probabilistic}.

The neural differentiable modeling framework provides an integrated learning platform where physics, represented by discretized PDE operators, is intricately fused with trainable neural network components to create advanced hybrid DL architectures, extending beyond conventional numerical solvers. Despite the promising strides made so far, this field is still in its early stages and requires further development. Specifically, the design of hybrid modeling architectures, which can be highly versatile and critical to learning performance, has not yet been fully explored and optimized. Further research is required to explore different architectural designs of neural differentiable models and to pinpoint potential training bottlenecks, thus pushing the boundaries of this novel hybrid learning paradigm in modeling complex turbulent flows. In this study, we introduce a data-driven surrogate model for predicting 2D turbulence using the differentiable neural modeling method. The proposed approach is structured into two key stages: (1) a hybrid differentiable neural solver is constructed to capture large-scale turbulent phenomena on a much coarser grid, and (2) a Bayesian conditional diffusion model is then developed to generate small-scale turbulent structures given the large-scale dynamics simulated by the hybrid solver. For the first stage, we explore and compare two innovative hybrid architecture designs for effectively and efficiently predicting large-scale turbulence. In one of our designs, the trainable neural networks are integrated within the differentiable PDE solver as correction terms, serving for SGS closures and truncation error corrections; the second hybrid design features a deeper fusion of trainable ML components and differentiable numerical PDEs, where discretization and interpolation schemes are parameterized as trainable neural networks. The proposed neural differentiable solver is compared against established methods: purely data-driven black-box models and conventional LES with physics-based SGS closures. Through comprehensive comparison and analysis, we aim to highlight the potential advantages in accuracy, computational efficiency, and versatile design offered by the neural differentiable modeling framework in the context of turbulence simulation.

\section{Methodology}
\label{sec:meth}
\subsection{Problem formulation} 
\label{sec:problem equation}

Two-dimensional (2D) turbulence, an idealized model for various large-scale geophysical and environmental flows influenced by rotation and/or stratification, provides fundamental insights into turbulent dynamics relevant to climate modeling, weather forecasting, and oceanography~\cite{mccomb2014homogeneous,guan2022stable, li2020fourier}. Characterized by chaotic spatialtemporal behavior, 2D Kolmogorov turbulence serves as an ideal testbed for developing mathematical and ML-based turbulence models~\cite{maulik2019subgrid,kochkov2021machine}. In this study, neural differentiable models are developed to capture the spatiotemporal dynamics of 2D Kolmogorov turbulence, which is governed by the incompressible Navier-Stokes (NS) equations, 
\begin{equation}
\label{eq:ns}
\begin{aligned}
\nabla\cdot\mathbf{u} = 0, \hspace{7em}&\mathbf{x}, t \in \Omega_f \times [0, T]\\
\frac{\partial{\mathbf{u}}}{\partial t} = -(\mathbf{u}\cdot\nabla){\mathbf{u}}+\nu\nabla^2\mathbf{u}-\frac{1}{\rho}\nabla{p} + \bm{f}, \hspace{3em} &\mathbf{x}, t \in \Omega_f \times [0, T]
\end{aligned} 
\end{equation}
where $t$ and $\mathbf{x} = \left[x,y\right]$ are time and Eulerian space coordinates, respectively; fluid velocity $\mathbf{u}(t, \mathbf{x})$ and pressure $p(t, \mathbf{x})$ are both spatiotemporal functions defined in $\Omega_f \subset \mathbb{R}^2$; $\rho$ and $\nu$ represent fluid density and kinematic viscosity of the fluid, respectively; $\bm{f}$ represents Kolmogorov forcing term, defined as~\cite{chandler2013invariant}, 
\begin{equation}
\label{eq:ns-force}
\begin{aligned}
\bm{f} = \chi \big[f_x, f_y\big] = \chi \big[\sin(k_f x),  0\big],
\end{aligned} 
\end{equation}
where $k_f = 5$ is the forcing wavenumber and $\chi = 1$ is the forcing magnitude. 
Solutions for velocity and pressure are uniquely determined by specified initial and boundary conditions (IC/BCs). However, numerically obtaining the solutions at all scales through DNS is computationally infeasible for most practical scenarios. 
\begin{figure}[ht!]
    \centering
    \includegraphics[width=1.0\textwidth]{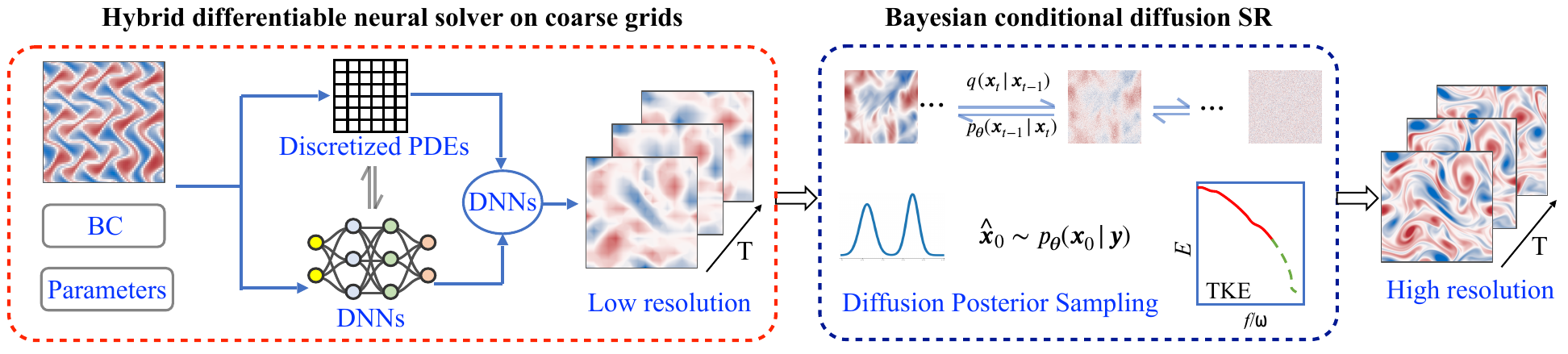}
    \caption{Overview of the hybrid neural differentiable modeling framework used for simulating spatiotemporal turbulence. It begins by taking the initial and boundary conditions, along with physical parameters, as inputs into a hybrid neural differentiable solver, which combines deep neural networks with coarse-grained PDE operators from governing equations to predict low-resolution turbulent flow fields. The outputs are refined into high-resolution ones by generating small-scale details using a Bayesian conditional diffusion model, enabling super-resolution (SR) predictions.}
    \label{fig:basic_idea}
\end{figure}
To address this challenge, we propose a hybrid data-driven neural differentiable modeling framework as illustrated in Fig.~\ref{fig:basic_idea}, where large-scale turbulent flows are captured using a hybrid neural differentiable solver on coarse spatiotemporal resolution, while small-scale turbulent structures are generated by an offline-trained Bayesian conditional diffusion model. Initially, the hybrid neural solver is constructed to accurately predict the flow solution at coarse scales. This is achieved by seamlessly integrating machine learning with physical laws through differentiable programming, operating on coarser grids to reduce computational load while preserving essential large-scale dynamics. Subsequently, a probabilistic diffusion model is trained to generate small-scale turbulent structures conditioned on the large-scale flow predictions. Using Bayesian posterior sampling, the conditional diffusion model functions as a zero-shot super-resolution network to recover high-wavenumber features in the second stage without retraining. This two-stage approach leverages the strengths of neural differentiable modeling and generative model to optimize computational efficiency, predictive accuracy, and model fidelity.

\subsection{Hybrid neural differentiable solver on coarse grids} 
\label{sec:designed_strategies}
In the first stage, a hybrid neural solver is developed on coarse grids to capture the large-scale dynamics by integrating a numerical solver with deep neural networks through differentiable programming. We propose and explore two innovative hybrid learning architectural designs. In Design I, trainable deep neural networks are integrated with the PDE solver, serving as both SGS closures and truncation error corrections. In Design II, the interpolation schemes of the PDE solvers are parameterized as trainable neural networks, learning advanced high-order discretization from data.  Details of these two architectural designs are provided below.

\subsubsection{Design I: hybrid neural-PDE correction model}
\label{sec:neuralcorrection}
\begin{figure}[tp!]
    \centering
    \includegraphics[width=1.0\textwidth]{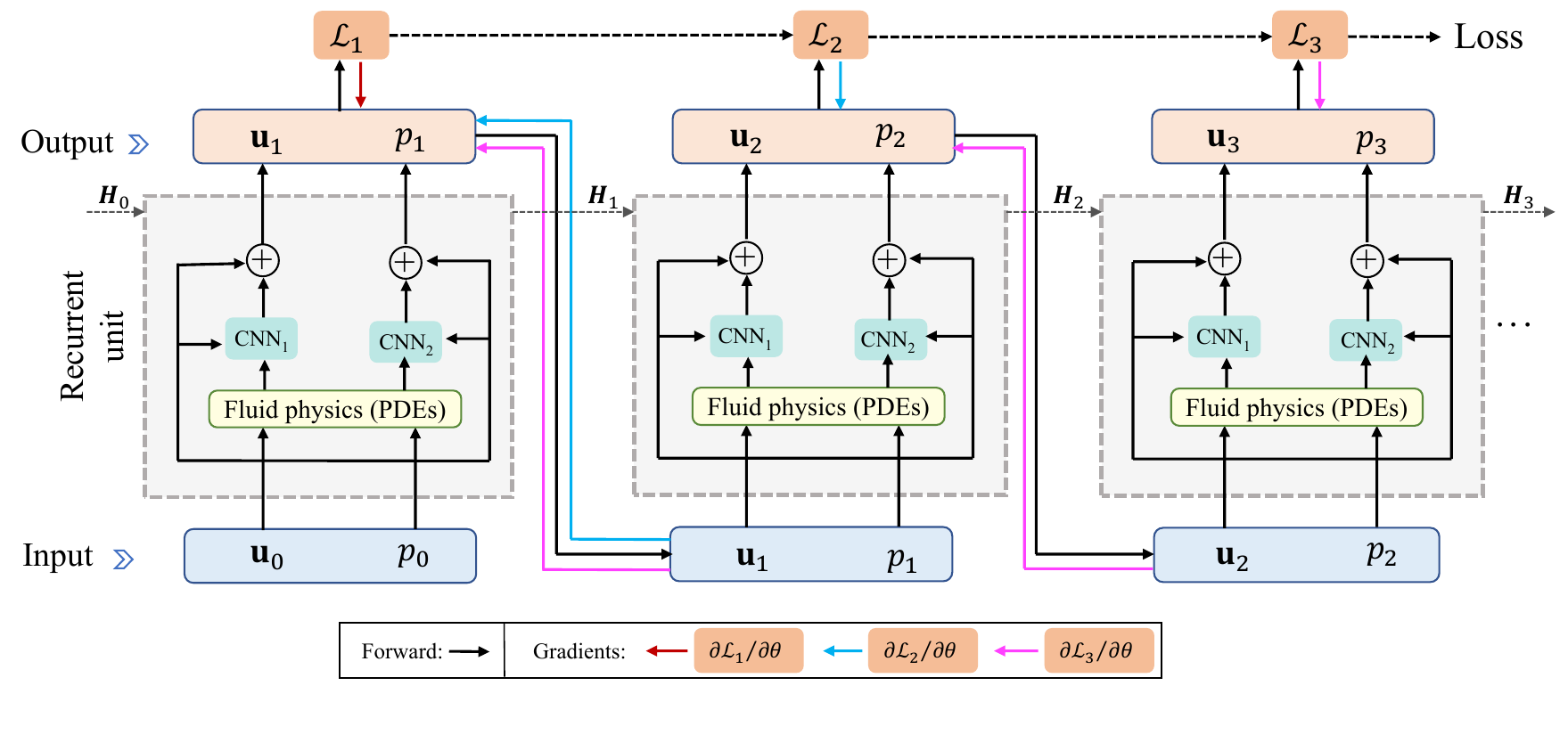}
    \caption{The schematic of the hybrid learning architecture I: neural-PDE correction model (NeuralPDE-Corr).}
    \label{fig:scheme_correction}
\end{figure}

In this design, we incorporate trainable DNN-based correction terms into the governing fluid equations to enhance the accuracy of flow predictions on coarse grids. As illustrated in Fig.~\ref{fig:scheme_correction}, the model adopts a recurrent neural network architecture, where the Navier-Stokes equations are encoded as fixed convolution operations, indicated as $\mathrm{ConvPDE}(\cdot)$. This is followed by a sequence of trainable convolutional LSTM (ConvLSTM) operators, $\mathrm{ConvLSTM}(\cdot)$ with trainable parameters $\bm{\theta}_{nn}$, which act as corrective adjustments to the flow predictions. Specifically, the flow solution $\bm{V}_t = [\mathbf{u}_t,p_t]^T$ at time $t$ is modeled as,
\begin{equation}
\bm{V}_t = \bm{V}_{t-1} + \mathrm{ConvPDE}(\bm{V}_{t-1}) + \mathrm{ConvLSTM}\Big(\bm{V}_{t-1}, \mathrm{ConvPDE}(\bm{V}_{t-1}), \bm{H}_{t-1}; \bm{\theta}_{nn}\Big),
\end{equation}
where $\bm{H}_{t-1}$ is the hidden state from previous step and the PDE-encoded convolution operation is defined as,
\begin{equation}
\mathrm{ConvPDE}(\bm{V}_{t-1}) = \Bigg\{
	\begin{aligned}
	& \delta\textbf{u}^*_t = \Delta t \Big(- (\textbf{u}_{t-1}\cdot\nabla)\textbf{u}_{t-1} + \nu\nabla^2\textbf{u}_{t-1} + \bm{f}\Big),\\
	& \delta p^*_t = S^{-1}\rho \nabla\cdot(\textbf{u}_{t-1} + \delta\textbf{u}^*_t)/\Delta t, 
	\end{aligned} 
\end{equation}
where $S$ is discretized Poisson operator and $\Delta t$ is time step. The detailed schematic of the recurrent network unit is depicted in Fig.~\ref{fig:details_correct}. In this setup, flow predictions -- including velocity and pressure from the previous step -- are advanced using the Navier-Stokes equations discretized on coarse grids, functioning as a coarse explicit fluid solver; the outputs are then processed through several convolutional layers, which are optimized during model training. The hybrid neural differentiable model is constructed and trained in an autoregressive manner, where the loss function is derived from the model predictions over multiple time steps. This setup requires every component of the hybrid model to be differentiable, enabling holistic end-to-end training. 
Conceptually, this learning architecture can be viewed as a coarse fluid solver coupled with trainable DNNs, which function as corrective mechanisms for unresolved subgrid-scale structures and numerical discretization errors. However, this model is beyond a mere combination of a classic coarse solver and DNN corrections, due to its sophisticated integration as a recurrent neural networks with hidden connections. We refer to this design as ``NeuralPDE-Corr''.

\subsubsection{Design II: Hybrid neural-PDE discretization model}
\label{sec:neural_scheme}

\begin{figure}[pt!]
    \centering
    \includegraphics[width=1.0\textwidth]{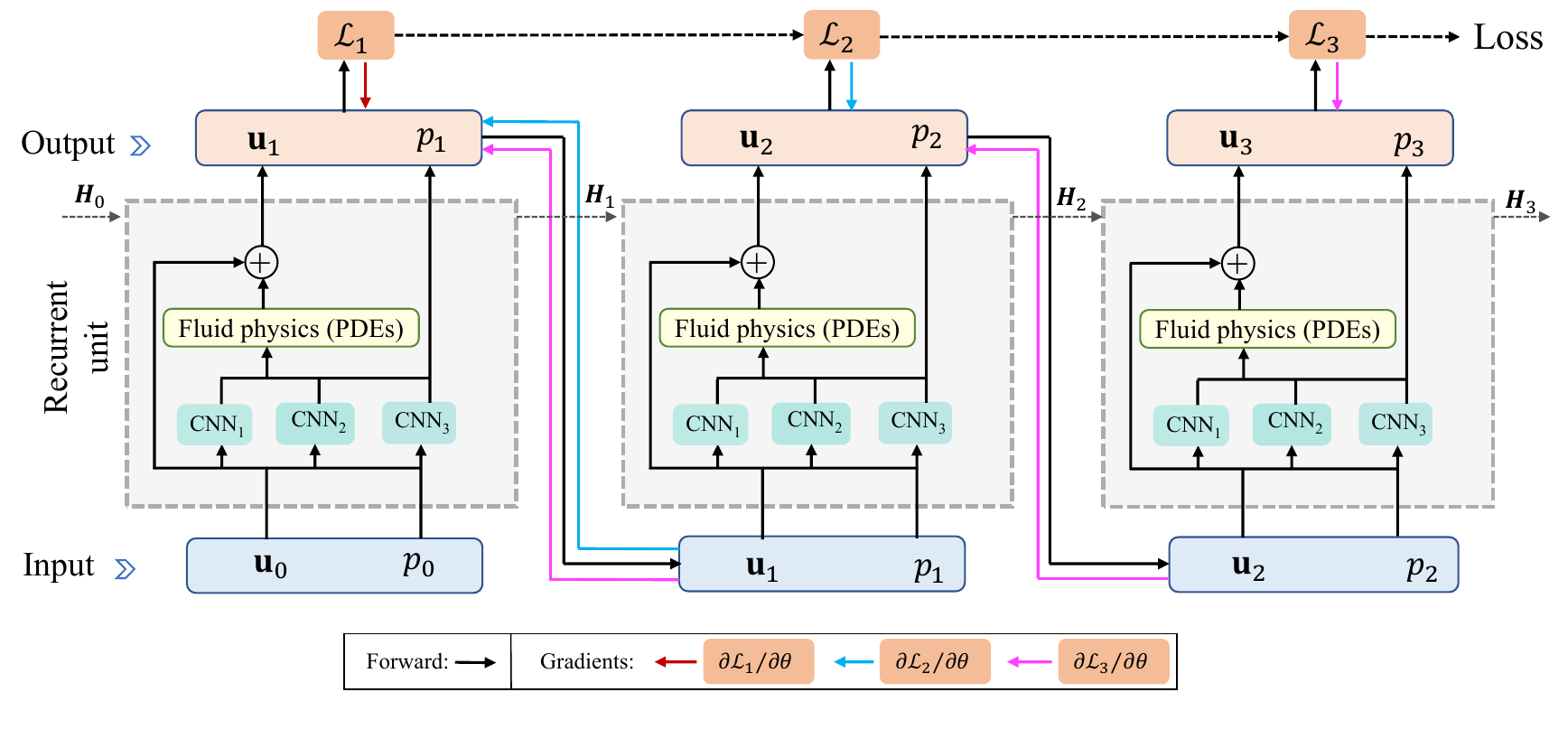}
    \caption{The schematic of the hybrid learning architecture II: neural-PDE discretization model (NeuralPDE-Discretize).}
    \label{fig:scheme_convection}
\end{figure}

In the second design, trainable DNNs are utilized to learn high-order numerical discretizations, facilitating effective discretized solution approximation at cell faces within a FVM framework to enhance the prediction accuracy with coarse grids~\cite{kochkov2021machine}. As depicted in Fig.~\ref{fig:scheme_convection}, the entire model operates as a recurrent neural network, where certain parts of the Navier-Stokes equations are formulated as fixed convolution operations, coupled with a series of trainable ConvLSTM layers. In contrast to Design I, the flow solution $\bm{V}_t = [\mathbf{u}_t,p_t]^T$ at time $t$ is modeled as,
\begin{equation}
\bm{V}_t = \Bigg\{
    \begin{aligned}
   &\bm{u}_{t-1} +  \mathrm{ConvPDE}\Big(\bm{u}_{t-1}, \bm{f}, \mathrm{ConvLSTM}_1(\bm{u}_{t-1}, \bm{H}_{t-1}; \bm{\theta}_{nn})\Big),\\
   &p_{t-1} +  \mathrm{ConvLSTM}_2\Big(p_{t-1}, \bm{H}_{t-1}; \bm{\theta}_{nn}\Big),  
    \end{aligned}
\end{equation}
where the PDE-encoded convolution operation is defined as,
\begin{equation}
\mathrm{ConvPDE}(\bm{u}_{t-1})= \delta\textbf{u}^*_t = \Delta t \Big(- (\textbf{u}^N_{t-1}\cdot\nabla)\textbf{u}^N_{t-1}+ (\nu\nabla^2\textbf{u}_{t-1}+\mathcal{D}_N)  + \bm{f}\Big)
\end{equation}
where $\textbf{u}^N_{t-1} = \mathrm{ConvLSTM}_1(\bm{u}_{t-1}, \bm{H}_{t-1}; \bm{\theta}_{nn})$ and $\mathcal{D}_N = \mathrm{ConvLSTM}_2(\nu\nabla^2\textbf{u}_{t-1}, \bm{H}_{t-1}; \bm{\theta}_{nn})$. 
Details of the recurrent unit for this design are presented in Fig.~\ref{fig:details_num}. In this model, the final predictions of velocity and pressure at time $t$ are advanced by the Navier-Stokes equations, discretized on coarse grids. Unlike the first design, which relies on DNNs for corrective and closure modeling, this approach uses trainable DNNs to learn high-order numerical discretization in a data-driven manner, replacing traditional second-order central interpolations typically used for convection at cell faces. Numerical dissipation and truncation errors, exacerbated by coarse grids, are compensated by the trainable neural networks. Moreover, to address the low solution accuracy of the Poisson equation on very coarse grids, the time stepping scheme for pressure is directly learned by neural networks. This integration of trainable neural networks with discretized PDEs is referred to as "NeuralPDE-Discretize." As with the first architecture, the NeuralPDE-Discretize model is meticulously constructed and systematically trained in an autoregressive manner, offering a more integrated and potentially complex approach. However, this extensive integration poses risks of numerical instability if neural network outputs are not adequately constrained~\cite{bar2019learning, zhu2019machine}. Comparative analyses in subsequent experiments will evaluate the performance of each design.


\subsection{Diffusion-based superresolution network for recovering small scales}
\label{sec:SR-NET}

Directly resolving all scales of turbulence, especially at high Reynolds numbers, is computationally infeasible. The proposed hybrid neural differentiable model overcomes this challenge by leveraging state-of-the-art (SOTA) generative AI techniques to recover high-frequency features associated with small scales. Specifically, we utilize probabilistic diffusion-based generative models, which have recently shown great promise in generating spatiotemporal turbulence~\cite{gao2023bayesian,du2024confild}. 
\begin{figure}[t!]
    \centering
    \includegraphics[width=\linewidth]{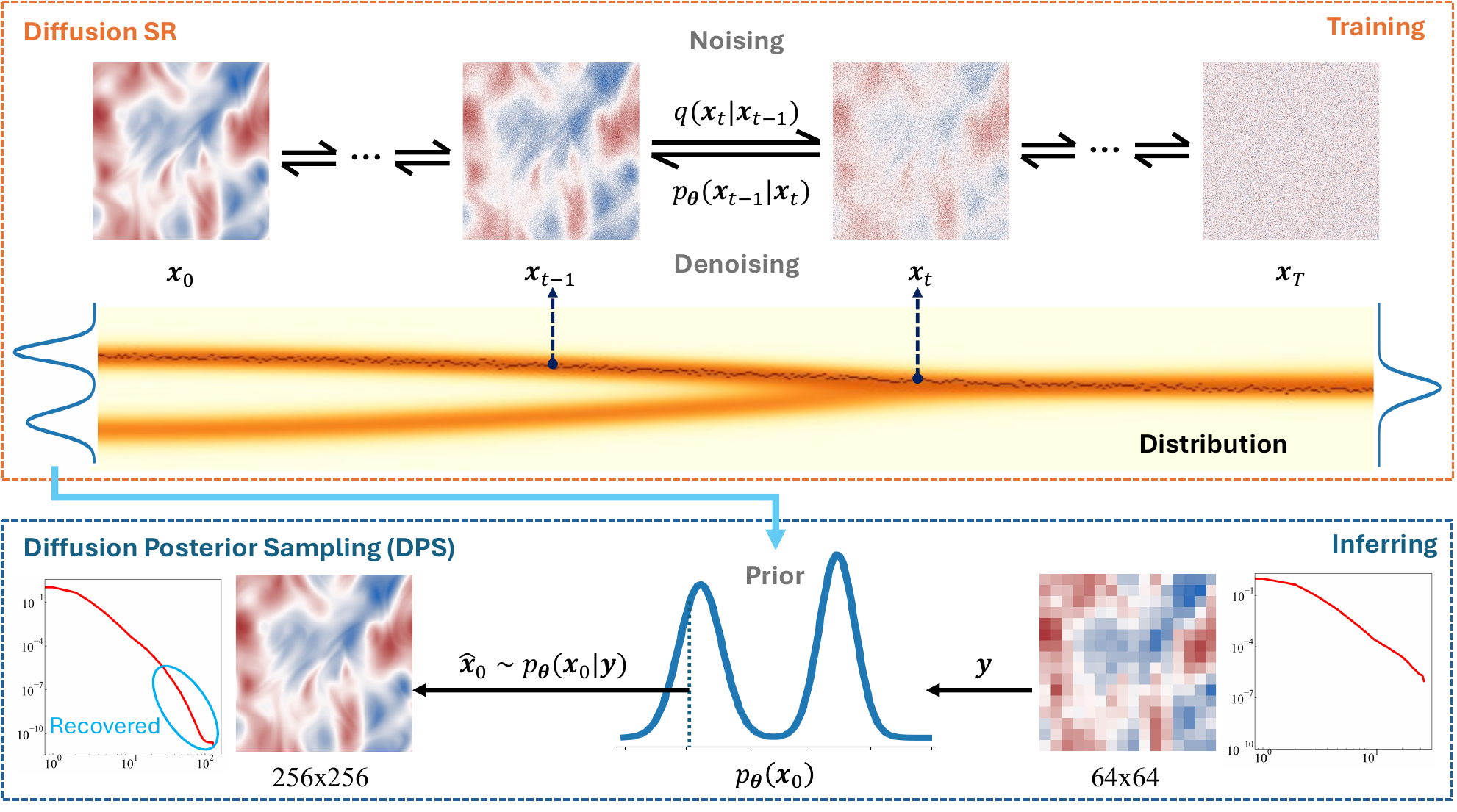}
    \caption{Schematics of the spectrum-decomposed diffusion-based SR model. Diffusion posterior sampling allows for conditional generation of high-resolution flow predictions for provided low-resolution flow solutions without retraining.}
    \label{fig:SR-DIffusion-DPS}
\end{figure}
As illustrated in Fig.~\ref{fig:basic_idea}, we propose a spectrum-decomposed diffusion-based super-resolution (SR) model, which is trained offline to enhance the low-resolution output from the coarse-grid hybrid neural differentiable solver by generating high-wavenumber turbulence. The diffusion model operates by learning a reverse Markovian diffusion process, which enables generation starting from a Gaussian random field and progressively converging to the target data distribution~\cite{dhariwal2021diffusion, ho2020denoising}. To achieve SR generation, a diffusion posterior sampling technique is used to enable conditional generation given the low-resolution coarse-grid flow predictions.  

During training, the high-resolution flow data $\mathbf{x}_0$ from the target distribution $q(\mathbf{x}_0)$ are sequentially corrupted based on a Gaussian transition kernel $q(\mathbf x_t|\mathbf x_{t-1}) = \mathcal{N}(\mathbf x_t ; \sqrt{1-\beta_t} \mathbf x_{t-1}, \beta_t I)$ parameterised by a noising schedule $\beta_t$, as shown in the upper section of Fig.~\ref{fig:SR-DIffusion-DPS}). This forward process involves sequentially adding Gaussian noise $\boldsymbol{\epsilon}$, such that $\mathbf x_{t+1} = \mathbf x_t + \beta_t \boldsymbol{\epsilon}, \quad \boldsymbol{\epsilon} \sim \mathcal{N}(0, I)$, leading to a forward Markov process, $\mathbf x_0 \rightarrow \mathbf x_1 \rightarrow \ldots \mathbf x_{t-1} \rightarrow \mathbf x_t \ldots \rightarrow \mathbf x_T$. The diffusion model learns to denoise the corrupted sample $\mathbf x_t \rightarrow \mathbf x_{t-1}$ by parameterizing the reverse Gaussian transition kernel using neural networks with trainable parameters $\bm{\theta}$ as,
\begin{equation}
   p_{\theta}(\mathbf x_{t-1}|\mathbf x_{t}) = \mathcal{N} \Big(x_{t-1} ; \frac{1}{\sqrt{1-\beta_t}} \Big(\mathbf x_t-\frac{\beta_t}{\sqrt{1-\bar\alpha_t}} \bm{\epsilon}_{\theta}(\mathbf x, t)\Big), \beta_t I\Big), \quad \bar\alpha_t = \prod_{s=1}^t (1-\beta_s) 
   \label{eq:reverse_gaussian}
\end{equation}
where $\boldsymbol{\epsilon}_{\theta}(\mathbf x, t)$ is a neural network approximating the noise $\boldsymbol{\epsilon}$ from $\mathbf x_t$. The spectrum-decomposed training process for $\boldsymbol{\epsilon}_{\theta}$ is detailed in Sec.~\ref{sec:training}. Once trained, the diffusion model is able to generate high-resolution flow samples from the learned distribution $\hat{\mathbf x}_0 \sim p_{\theta}(\mathbf x_0)$, which approximates the true distribution $q(\mathbf x_0)$, underlying the training data. These distributions are referred to as prior distributions.

For SR generation given low-resolution flow predictions $\mathbf{y}$ from the coarse-grid hybrid neural solver, the diffusion posterior sampling (DPS) method~\cite{chung2022diffusion} is utilized. As depicted in the lower part of Fig.~\ref{fig:SR-DIffusion-DPS}, for a given mapping function $\mathcal{M}: \mathbf{x}_0 \mapsto \mathbf{y}$, we can sample the trained diffusion model conditioned on low-resolution flow predictions $\mathbf{y}$ without retraining. This is achieved by modifying the sampling process of the unconditionally trained diffusion model $q_{\theta}(\mathbf{x}_0)$ (i.e., prior) with a updated score function using the gradient of the log likelihood function $-\nabla_{\mathbf{x}_i}||\mathbf{y}-\mathcal{M}(\hat{\mathbf{x}}_0)||^2_2$. More details can be found in Du et al.~\cite{du2024confild}.

\subsection{Training details of the entire framework}
\label{sec:training}

This subsection offers an overview of the training recipe, which largely influences the performance of the proposed model. Understanding and optimizing these training details are crucial to leveraging the full potential of the hybrid neural differentiable model in learning turbulent flows. Specifically, the training of low-resolution hybrid neural solver and diffusion-based SR module are decoupled in two stages. The source of our training and testing data is high-resolution DNS performed on a grid of $N_x \times N_y = 2048 \times 2048$, utilizing a fractional step method as detailed in \cite{fan2024differentiable}. For training the coarse grid neural solver, the high-resolution DNS data $\bm{\Phi}^{\mathrm{DNS}}$ undergoes filtering and down-sampling to a coarse resolution of $64 \times 64$, forming the low-resolution dataset, $\mathcal{A}_\mathrm{train}= \{\bm{\tilde{\Phi}_i}\}_{i=1}^{1600}$, where $\bm{\tilde{\Phi}} = [\tilde{\Phi}^1, \tilde{\Phi}^2, \tilde{\Phi}^3]^T$ corresponds to three distinct initial conditions.

\paragraph{Training for coarse-grid neural solver} 
Both hybrid Neural-PDE models are trained autoregressively in a sequence-to-sequence (Seq2Seq) manner, where the loss function is defined as follows,
\begin{equation}
\mathcal{L}(\boldsymbol{\theta}) =\alpha_1 \frac{1}{N} \sum_{t=0}^{N-1} \Big\lVert \tilde{\mathbf{u}}^{t}(\mathbf{x};\bm{\theta}) - \mathbf{u}_d^{t}(\mathbf{x}) \Big\rVert_{L_2}^{2} +\alpha_2 \frac{1}{N} \sum_{t=0}^{N-1} \Big\lVert \tilde{p}^{t}(\mathbf{x};\bm{\theta}) - p_d^{t}(\mathbf{x}) \Big\rVert_{L_2}^{2}
\label{eq:loss_component}
\end{equation}
where $\lVert \cdot \rVert_{L_2}$ represents the L2 norm, $N$ indicates the total number of rollout time steps, $\tilde{\Box}$ denotes the rollout predictions, and the subscript ${\Box}_d$ indicates the labeled data. The specified weights, $\alpha_1=1$ and $\alpha_2=100$, are based on the relative importance of velocity and pressure magnitudes. This structured loss is minimized over a long rollout trajectory using stochastic gradient descent, enhancing both the accuracy and stability of the model's autoregressive inference performance as noted in~\cite{um2020solver, kochkov2021machine}. This multi-step rollout training process is enabled by differentiable programming, which allows the gradient to be back-propagated over the entire rollout trajectory. However, implementing Seq2Seq training through differentiable programming may introduce specific challenges: (1) Training instability: the hybrid neural solver is prone to training instability, especially in the early training phase. This is because the initialization of trainable neural network parameters is nonphysical, which can result in numerical instability in the PDE-encoded portion and lead to training failure. (2) Extensive memory demands: training over a long trajectory require substantial GPU memory, since the gradients of nested functions from numerous model rollout steps have to be stored for end-to-end optimization. To address these challenges effectively, we employ the following training recipe: (i) Training is structured with a total of $N=200$ unrolling steps, with the loss being calculated every 2 steps. The learning step, $dt_{\text{neural}}$, is defined as $8 \times dt_{\text{physics}}$, matching the time step used in the training dataset $\mathcal{A}_{\text{train}}$. (ii) We gradually increase the number of unrolling steps from $N=40$ to $N=200$ in increments of $\delta=40$ through transfer learning, which helps in maintaining stability over long trajectories. (iii) To circumvent memory limitations, training is parallelized across multiple GPUs. The differentiable neural solver is developed in JAX, taking advantage of jax.pmap to parallelize the computations across batch dimensions. Currently, four batches with differing initial conditions are used, and the gradients from each GPU are aggregated to update the training parameters each epoch. In the hybrid neural models, all CNN blocks are uniformly configured with five layers, following the channel configuration of $[32, 64, 64, 32, 1]$ with a trainable $3 \times 3$ convolutional kernel. A Leaky rectified linear unit (LeakyReLU) serves as the activation function for all layers except the last one, which is linear. The optimization settings are as follows: (a) Initial Learning Rate: $10^{-4}$, (b) Optimizer: Adam, (c) Scheduler: Cosine Decay Schedule with alpha=$10^{-9}$. 

\paragraph{Training for spectrum-decomposed diffusion-based SR model}
The training strategy for the diffusion-based SR portion is illustrated in Fig.~\ref{fig:SR-DIffusion-DPS}. The diffusion model is trained unconditionally to learns the prior distribution $q_{\theta}(\bm{x}_0)$ from the high-resolution DNS dataset. During inference, it adapts to conditionally generate small-scale features based on the large-scale flow predictions. To enhance the model's capability of generating high-wavenumber flow structures, we spectrally decompose the high-resolution DNS data, $\bm{\Phi}^{\mathrm{DNS}}$, into two components: $\bm{\tilde{\Phi}}_{\mathrm{H}}$ and $\bm{\tilde{\Phi}}_{\mathrm{L}}$, representing two different energy scales. The decomposition is based on the cutoff wavenumber $k_{\mathrm{cut}}$, which are determined by the spatial resolution of the coarse-grid neural solver. This spectral decomposition forms the training dataset $\mathcal{B}_{\mathrm{train}} = [\bm{\tilde{\Phi}}_{\mathrm{H}}, \bm{\tilde{\Phi}}_{\mathrm{L}}]^T$, allowing for focused learning on distinct wavenumber bands.

Our spectrum-decomposed diffusion-SR models are trained on $\mathcal{B}_{\mathrm{train}}$, and generation are performed from two separate channels corresponding the two distinct wavenumber bands, which are combined to for the generated high-resolution flow solution. Separate diffusion models are trained for each velocity component, $\bm{u} = [u, v]^T$, and pressure, $p$, focusing solely on the high-resolution dataset to ensure that the fine-scale features are precisely learned without any low-resolution input. The diffusion-SR models utilize a U-net architecture to approximate the noise component $\boldsymbol{\epsilon}_{\theta}$, through the optimization of the following objective function~\cite{dhariwal2021diffusion}, 
\begin{equation}
  L_{simple} (\theta) := \mathbb{E}_{t,\mathbf{x}_0, \boldsymbol{\epsilon}} \Big[ || \boldsymbol{\epsilon} - \boldsymbol{\epsilon}_{\theta}(\sqrt{\bar\alpha_t}\mathbf{x}_0 + \sqrt{1-\bar\alpha_t}\boldsymbol{\epsilon}, t) ||^2 \Big]  
\end{equation}
This spectrum-decomposed training strategy not only enhances the model’s performance in terms of predicting small-scale flow features but also ensures stability and efficiency in handling extensive data scales, which effectively addresses the computational challenges posed by high-resolution data handling and long training trajectories.

\subsection{Baseline models for comparison}
In evaluating the effectiveness of the proposed hybrid neural models, it is essential to compare their performance against established modeling approaches. Two baseline models, one purely data-driven and the other purely physics-based, are utilized for this purpose. These two baseline models represent conventional approaches in ML and CFD, serving as reference points to study the performance of the proposed hybrid learning strategy.

\subsubsection{Purely data-driven neural solver}
\label{sec:blackbox}
To evaluate the role of integrating discretized flow physics into the neural solver, we compare our hybrid model against a purely data-driven neural solver that employs a similar ConvLSTM architecture but lacks any explicit incorporation of physical laws. This ``black-box'' approach utilizes the ConvLSTM to predict future flow states solely based on historical data, without any physics-based guidance. The recurrent network can be formulated as,
\begin{equation}
\bm{V}_t = \bm{V}_{t-1} + \mathrm{ConvLSTM}\Big(\bm{V}_{t-1},  \bm{H}_{t-1}; \bm{\theta}_{nn}\Big),
\end{equation}
which illustrates how this model leverages sequential data to autoregressively update velocity and pressure fields at each time step. The primary distinction from the hybrid models shown in Figs.~\ref{fig:scheme_correction} and~\ref{fig:scheme_convection} lies in the absence of any fluid physics modules, highlighting the model's reliance on the learned patterns purely from data rather than governing equations. 
\begin{figure}[H]
    \centering
    \includegraphics[width=1.0\textwidth]{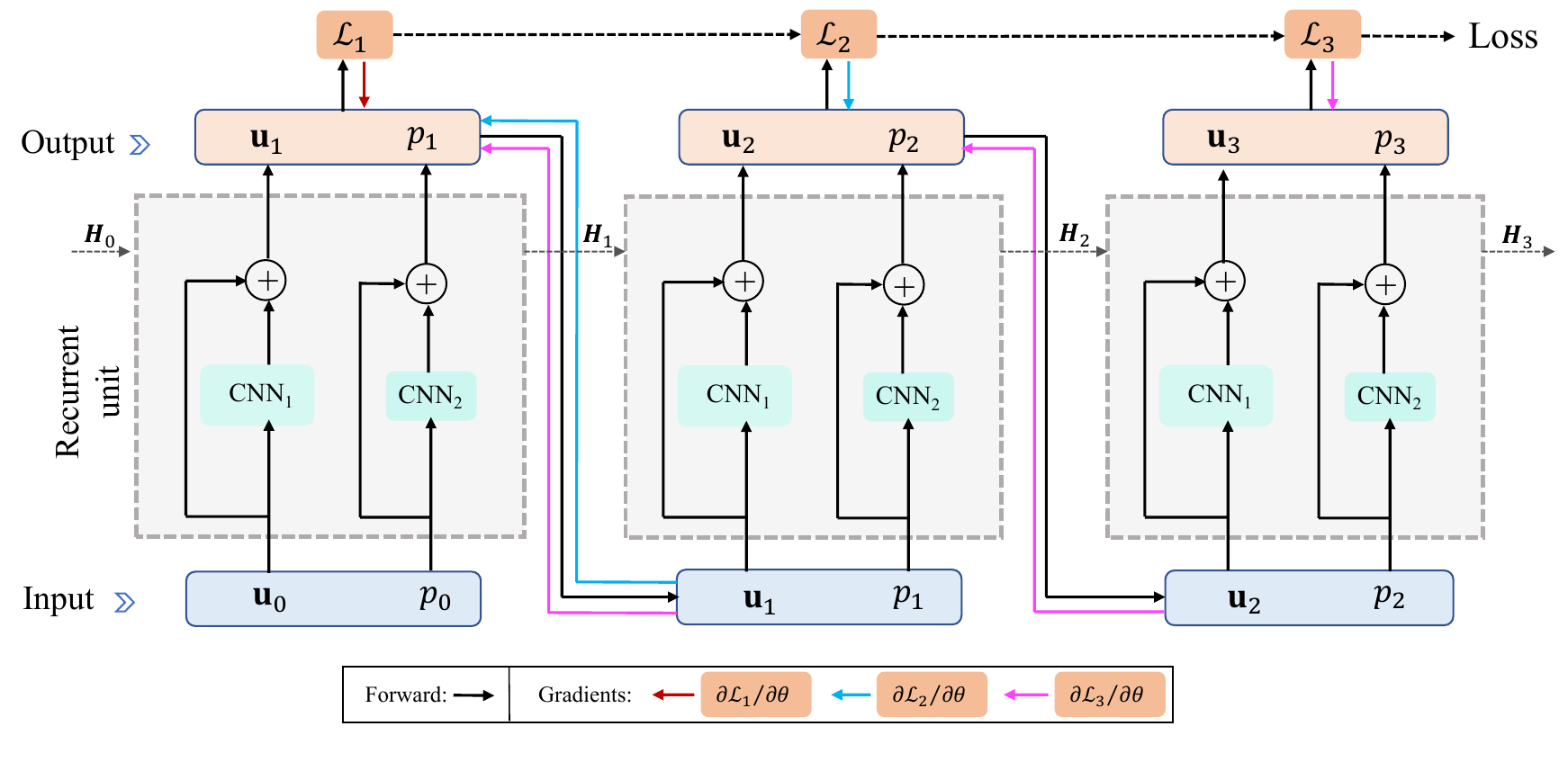}
    \caption{Schematic of the purely data-driven model illustrating how velocity and pressure are predicted using $\mathrm{CNN}_1$ and $\mathrm{CNN}_2$. The model is trained and applied in an autoregressive manner.}
    \label{fig:scheme_blackbox}
\end{figure}



\subsubsection{Purely physics-based solver with classic SGS closures}
\label{sec:sgs}
\begin{figure}[H]
    \centering
    \includegraphics[width=1.0\textwidth]{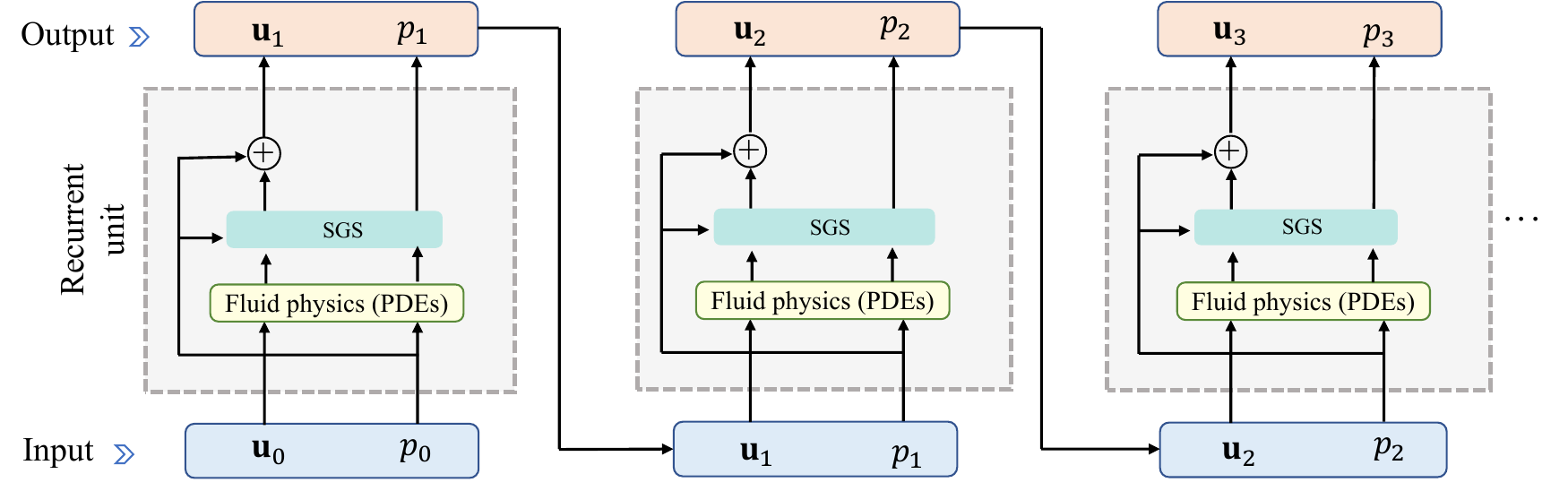}
    \caption{Schematic of the purely physics-based solver with classic SGS closures.}
    \label{fig:LES}
\end{figure}

In our comparison study, we also include a purely physics-based solver utilizing
classic SGS closures, commonly known as LES. The LES model serves as a benchmark for evaluating the benefits of incorporating learned elements via DNNs against traditional numerical methods that rely entirely on established governing physics. Specifically, the well-established Smagorinsky SGS model is utilized to address unresolved physics due to coarse computational grids. Unlike data-driven models, the Smagorinsky model provides corrective terms by approximating the effects of smaller, unresolved scales through the calculation of turbulent viscosity based on local strain rates. This method offers a physics-derived approach for representing the dissipative behavior of the scales not captured on coarse grids. The filtered Navier-Stokes equations, which incorporate the SGS model, are expressed as follows,
\begin{equation}
\label{eq:ns-sgs}
\frac{\partial{\mathbf{u}}}{\partial t} = -(\mathbf{u}\cdot\nabla){\mathbf{u}}+\nu\nabla^2\mathbf{u}-\frac{1}{\rho}\nabla{p} + \mathbf{f} - \nabla \cdot \bm{\tau}^{SGS}
\end{equation}
where $\bm{\tau}^{SGS}=-2\nu_{SGS}\Bar{\mathbf{S}}$ is the subgrid-scale stress tensor, $\Bar{\mathbf{S}}$ is the strain rate and the SGS viscosity $\nu_{SGS}$ is given by Smagorinsky model:
\begin{equation}
\label{eq:sgs-vis}
\nu_{SGS}=(C_s \Delta)^2\left|\bar{S}\right|
\end{equation}
where Smagorinsky constant is $C_s=0.2$, $\Delta$ is the grid size. In essence, the traditional LES model operates similarly to a nono-trainable recurrent neural network. As illustrated in Fig.~\ref{fig:LES}, it systematically utilizes physical laws to define convolution operations that compute each new state based on the solutions at previous time step, mirroring the standard recurrent network structure without hidden memory and learning capability. Namely, the LES model can be mathematically expressed as an autoregressive recurrent network,
\begin{equation}
\bm{V}_t = \bm{V}_{t-1} + \mathrm{ConvPDE}(\bm{V}_{t-1}) + \mathrm{SGS}(\bm{V}_{t-1}),
\end{equation}


\section{Numerical results}
\label{sec:results}

\subsection{Predictions at low resolution by hybrid neural models}
\label{sec: comparative analysis}
\subsubsection{Temporal forecasting starting from training ICs}
\begin{figure}[htp!]
    \centering
    \includegraphics[width=0.9\textwidth]{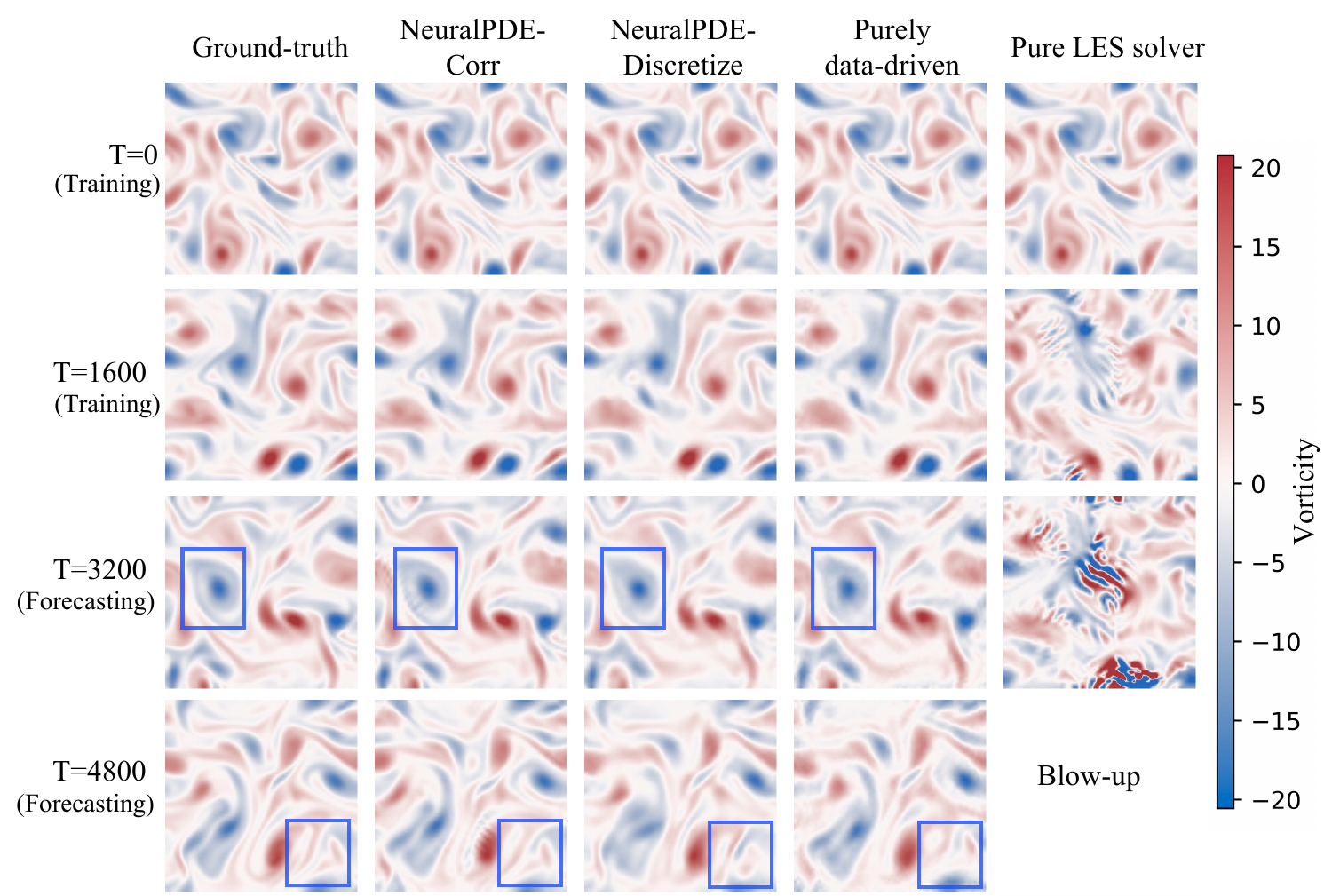}
    \caption{Temporal progression of vorticity contours predicted by different models on a coarse mesh, presented at key stages from training through extended forecasting. Notably, blue boxes highlight specific instances in the forecasting phase where significant differences between the models become apparent, including a blow-up scenario observed in the purely physics-based LES solver at $T=4800$.}
    \label{fig:HIT_contour_vort}
\end{figure}

The trained hybrid neural models were applied to forecast the spatiotemporal dynamics of Kolmogorov turbulence, beginning from the initial conditions used during training. To facilitate the comparison across different models, the superresolution components were excluded in this analysis. All results were generated using a coarse mesh of $64 \times 64$. The vorticity contours for each model are presented in Fig.~\ref{fig:HIT_contour_vort}, spanning both the training window from $T=0$ to $T=1600$ and the forecasting window from $T=1600$ to $T=4800$. 
Overall, both hybrid neural models, NeuralPDE-Corr and NeuralPDE-Discretize, successfully forecast the dynamics over a temporal trajectory three times longer than their training duration. However, there are still noticeable differences in their performance. The NeuralPDE-Corr model exhibits slight nonphysical oscillations in the vorticity predictions, particularly visible at $T = 3200$ and persisting through $T = 4800$, as highlighted by the blue frames. In contrast, NeuralPDE-Discretize model manages to eliminate these spurious oscillations, although the predicted patterns has a larger discrepancy in some regions after extensive rollouts. These performance differences can be attributed to the architectural differences between the two hybrid models. The NeuralPDE-Corr model, as detailed in Section~\ref{sec:neuralcorrection}, incorporates trainable convolutional layers as the final output layers, with each step's outputs directly produced by the CNNs. This configuration implies that while the predicted velocity and pressure fields are informed by the underlying physical laws, they are not strictly constrained by them, potentially leading to slight nonphysical oscillations. Conversely, the NeuralPDE-Discretize model, described in Section~\ref{sec:neural_scheme}, begins with trainable CNNs, which are seamlessly integrated with the physics by forming the discretized governing PDEs. This effectively transform the model into a FV-based solver, where traditional interpolation schemes are replaced by trainable neural networks. This integration ensures that outputs adhere closely to discretized physical laws. However, due to the chaotic nature of turbulence, even minor deviations in flux predictions can lead to significantly different patterns over an extended rollout trajectory, as evidenced by the differences observed at $T=4800$.

\begin{figure}[pt!]
    \centering
    \includegraphics[width=0.9\textwidth]{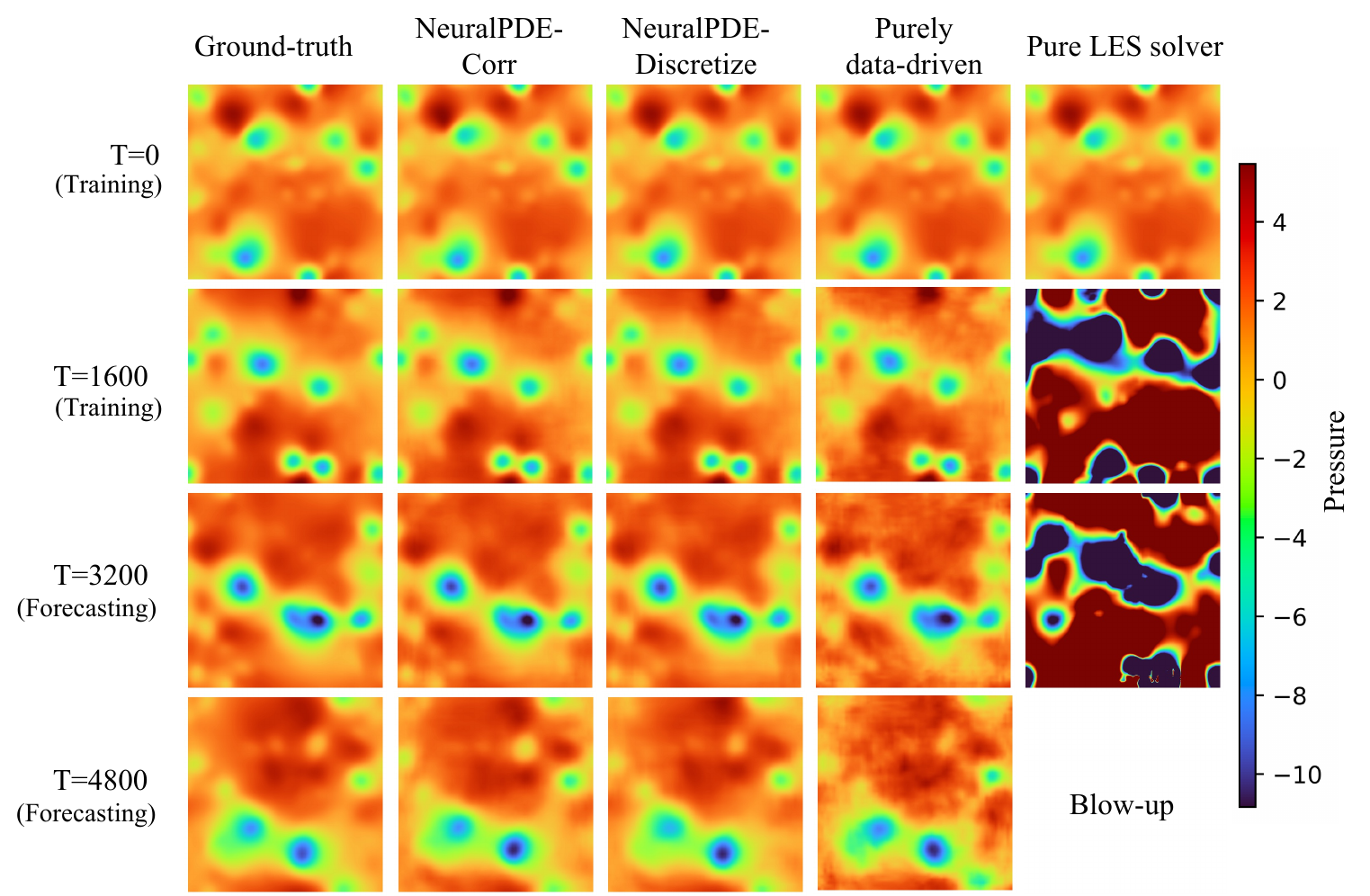}
    \caption{Pressure predictions from different models on a coarse mesh.}
    \label{fig:HIT_contour_pre}
\end{figure}
The forecasts from two baseline models, one purely data-driven and the other entirely physics-based, are compared. The purely data-driven model, which relies solely on historical data without incorporating physical laws, demonstrates competent performance in temporal forecasting from training initial conditions. However, it still slightly underperforms compared to the hybrid models due to its tendency to exhibit vorticity diffusion over extended simulations, leading to blurred vortex boundaries, particularly among larger vortices, as shown in Fig.~\ref{fig:HIT_contour_vort}. Additionally, significant nonphysical oscillations and non-smoothness are apparent in the predicted pressure fields, as shown in Fig.~\ref{fig:HIT_contour_pre}. While it performs well with known initial conditions, the purely data-driven approach’s effectiveness significantly declines when faced with unseen initial conditions, highlighting its limitations in generalizing beyond the training scenarios, which will be further discussed in the subsequent subsection. As for the purely physics-based baseline model, i.e., the traditional LES model with the Smagorinsky SGS closure, it fails to maintain stability on a coarse computational mesh and experiences a blow-up by $T = 3200$. Despite the Courant–Friedrichs–Lewy (CFL) number being $0.64$, numerical instability occurs due to inadequate grid resolution and the use of a basic explicit time integration scheme. This comparison underscores the superiority of hybrid neural models that integrate ML with physics-based solvers, enabling them to operate effectively on much coarser grids and with larger time steps compared to conventional numerical solvers.

\begin{figure}[pt!]
    \centering
    \includegraphics[width=0.7\textwidth]{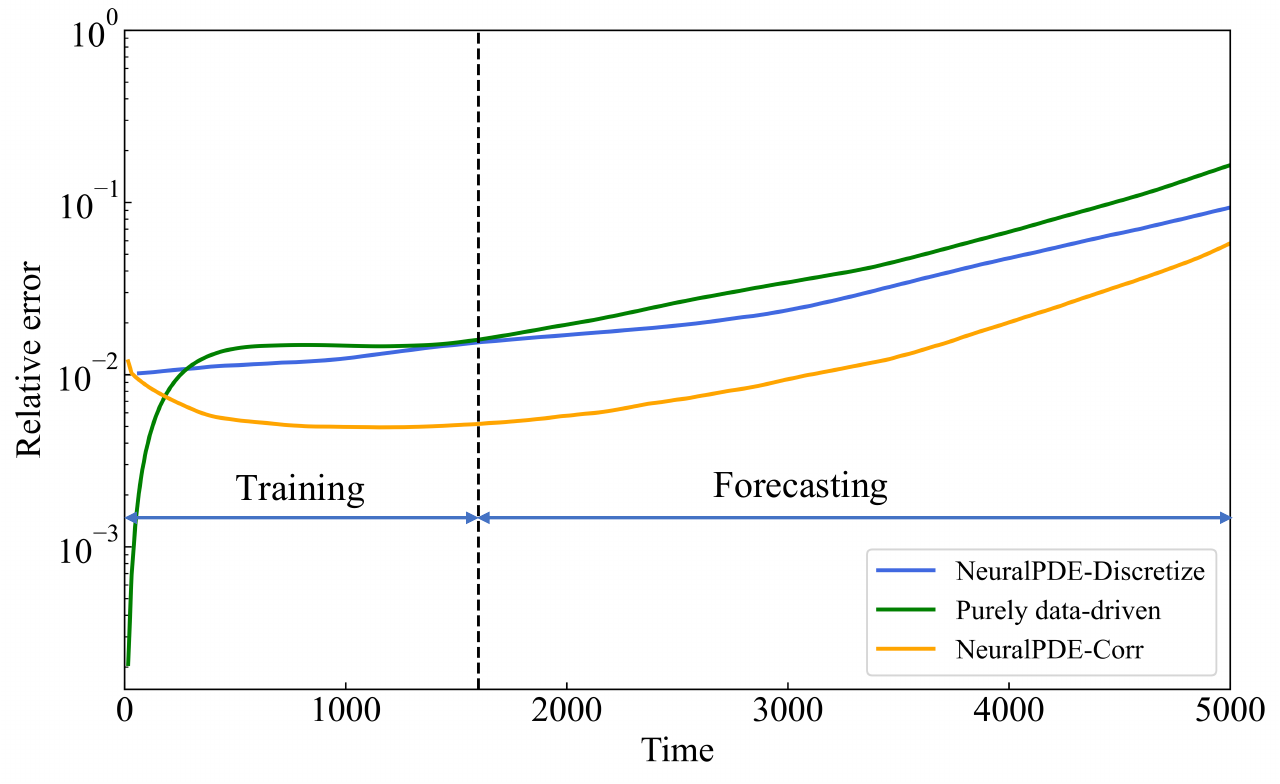}
    \caption{Relative error comparison across all models for initial conditions seen during training. Due to its substantially higher error, the pure LES solver's results is excluded. } 
    \label{fig:error}
\end{figure}
To further quantitatively assess the error propagation in time for all models, the relative errors $\epsilon^t$ of model prediction at each time step $t$ is computed as follows, 
\begin{equation}
  \epsilon^t=\frac{1}{N_\lambda} \sum_{i=0}^{N_{\lambda}}  \left(\frac{\lVert \mathbf{\widetilde{u}}^t(\lambda_i)-\mathbf{u}^t_d(\lambda_i) \rVert _{2}}{\lVert \mathbf{u}^t_d(\lambda_i) \rVert _{2}} \right) 
  \label{eq:relative error}
\end{equation}
where $N_\lambda$ is the size of parameter ensemble. The computed relative errors for each model, shown in Fig.~\ref{fig:error}, reveal that the hybrid neural models, NeuralPDE-Corr and NeuralPDE-Discretize, generally outperform the purely data-driven model throughout the forecasting phase. Notably, the error trajectory of the purely data-driven model (green line) starts low but guickly grows by an order of magnitude during the forecast period, a common issue in data-driven forecasting due to error accumulation. Of the two hybrid models, NeuralPDE-Corr (orange line) consistently exhibits lower errors than NeuralPDE-Discretize (blue line). This can be attributed to the different operational mechanics of the two hybrid architectural designs. As indicated by the loss defined in Eq.~\ref{eq:loss_component}, both models aim to minimize the data mismatch in terms of instantaneous flow patterns. This objective is more straightforwardly achieved for the NeuralPDE-Corr model, which employs neural networks to refine outputs from the coarse PDE solver, effectively replicating the flow patterns for the training scenario. In contrast, NeuralPDE-Discretize, which learns interpolation schemes by ML within the numerical solver, is closer to a classic numerical solver, which can simulate different flow patterns after long-term model rollouts given the chaotic nature of turbulence. This difference becomes particularly pronounced in the forecasting phase, where the inherently chaotic nature of turbulence leads to divergent outcomes from similar initial conditions. Consequently, while both models aim to match instantaneous flow patterns, NeuralPDE-Corr achieves a more accurate alignment with the data when evaluated by the relative MSE.


\begin{figure}[pt!]
    \centering
    \includegraphics[width=0.6\textwidth]{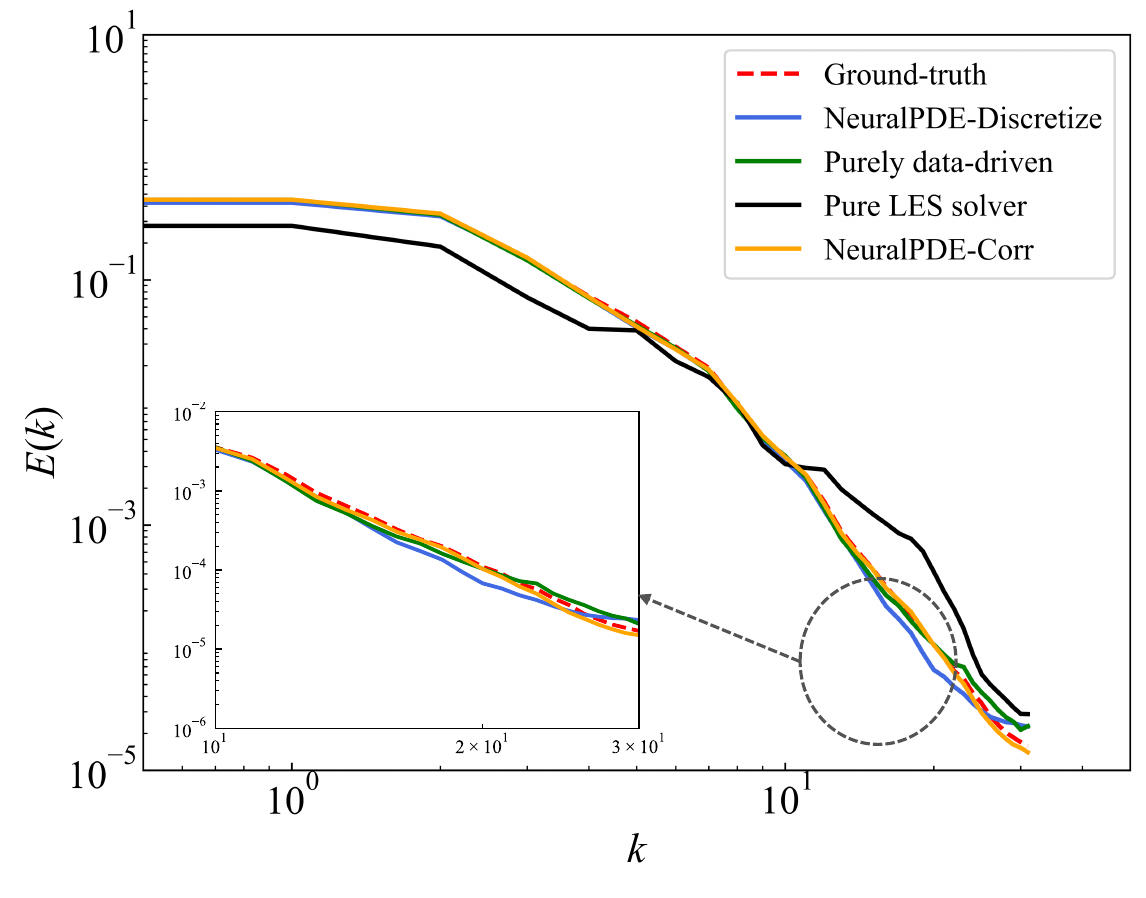}
    \caption{Energy spectra comparison across different models, averaged over the period from $T=2400$ to $T=4800$ to ensure decorrelation from initial conditions. For the Pure LES solver, energy spectra are calculated from $T=2400$ to $T=3200$ only, due to model divergence beyond $T=3200$.}
    \label{fig:train_energy}
\end{figure}
Fig.~\ref{fig:train_energy} presents the comparison of energy spectra across different models, benchmarked against ground truth, obtained from high-resolution DNS with spectral filtering (red line). Overall, the LES on very coarse space-time resolution fails to capture the energy spectrum, while the turbulence predictions from both hybrid and purely data-driven models align well with the ground truth, demonstrating their robustness in capturing essential statistical features by incorporating data. From the zoomed-in view, slight discrepancies in the high wave number regions ($k>10$) can be observed in the predictions by the NeuralPDE-Discretize and purely data-driven models, with the former underpredicting and the latter overpredicting energy at higher wave numbers. In contrast, the NeuralPDE-Corr model closely matches the filtered DNS statistics, precisely replicating the energy distribution. 


\subsubsection{Temporal forecasting from randomly generated ICs}
\label{sec: testing}
The inherently chaotic nature of turbulence leads to diverse flow patterns from different initial conditions. To assess the generalizability of our trained models, we conducted tests using initial conditions that were not seen during the training phase. These initial velocity and pressure fields were synthesized by sampling a Gaussian process formulated through Karhunen-Loeve (KL) expansion~\cite{gao2021bi}. Each model was then tasked with simulating the flow over a period of $T=3200$, identical to the number of inference steps used in earlier scenarios.
\begin{figure}[htp!]
    \centering
    \includegraphics[width=0.9\textwidth]{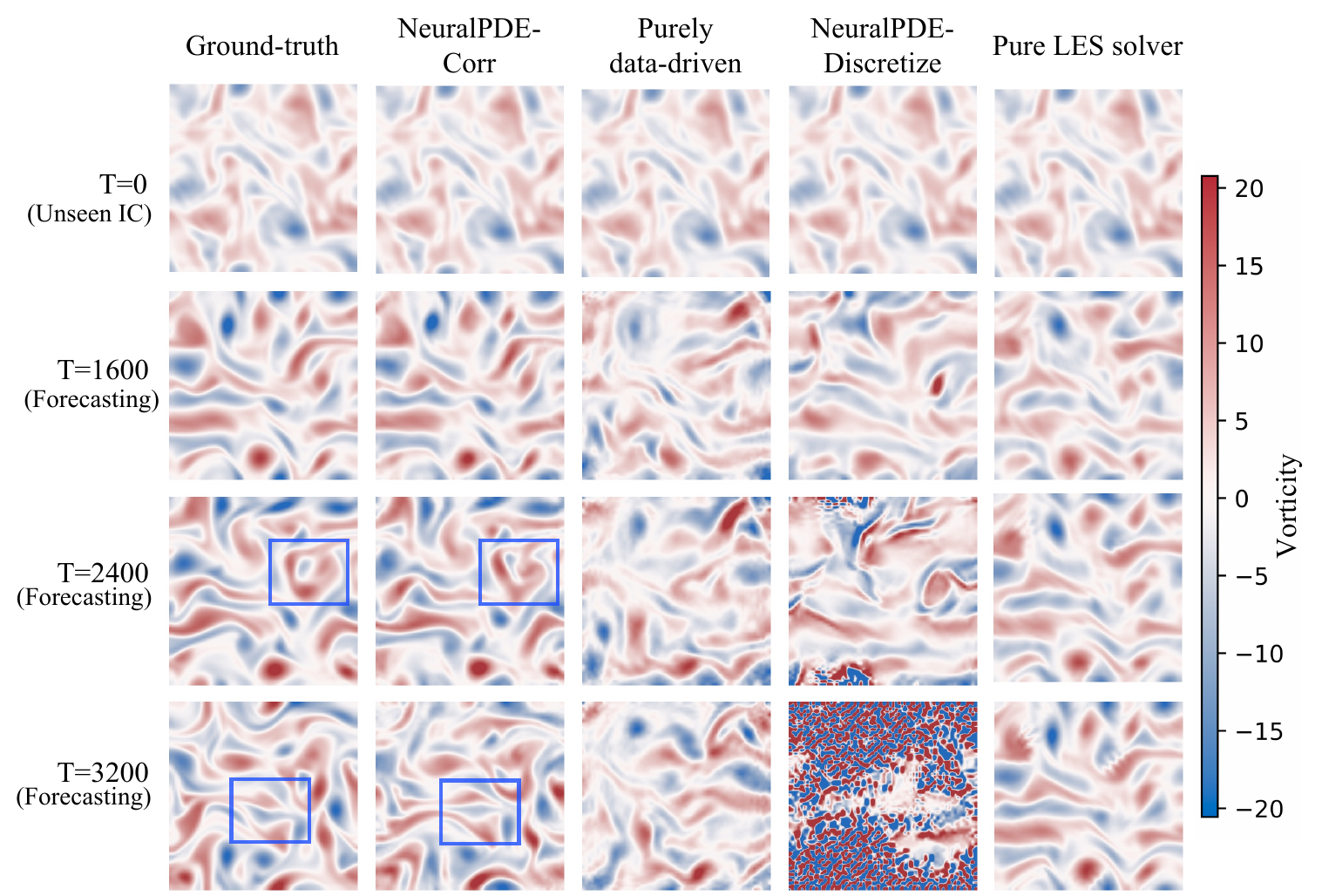}
    \caption{Temporal progression of vorticity contours predicted by different models on a coarse mesh, starting from one randomly generated initial conditions which has not been seen during training.}
    \label{fig:vorticity_test}
\end{figure}
\begin{figure}[t!]
    \centering
    \includegraphics[width=0.9\textwidth]{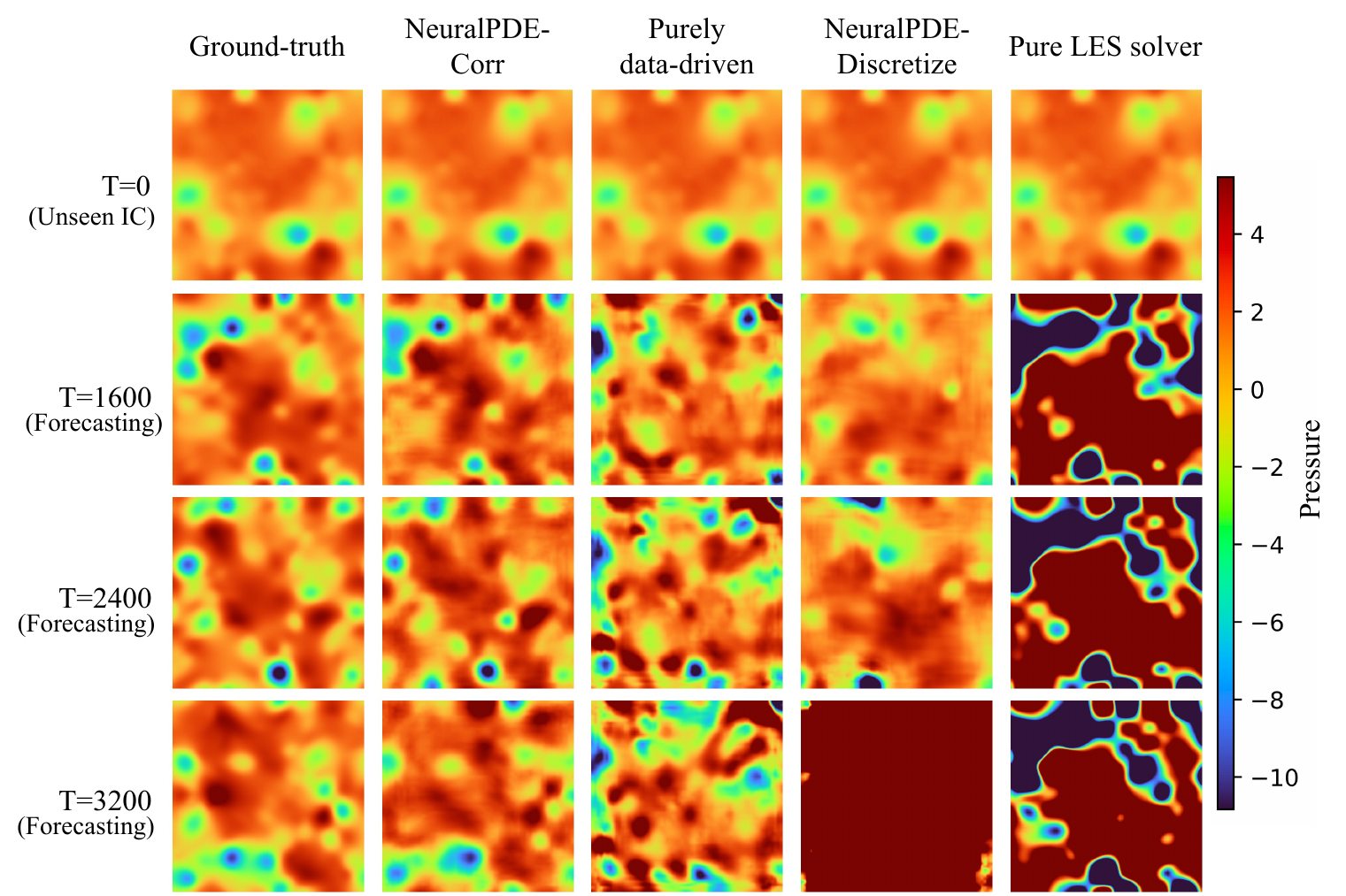}
    \caption{Temporal progression of pressure contours predicted by different models on a coarse mesh, starting from one randomly generated initial conditions which has not been seen during training.}
    \label{fig:pressure_test}
\end{figure}
As observed in the vorticity and pressure contours presented in Fig.~\ref{fig:vorticity_test} and Fig.~\ref{fig:pressure_test}, only the NeuralPDE-Corr model remains closely aligned with the ground-truth after a long-term rollout from unseen initial conditions. It effectively captures the dynamics of Kolmogorov turbulence with only minor deviations, which do not detract from the model's overall accuracy in replicating the turbulence dynamics. In stark contrast, the purely data-driven model fails to generalize under these unseen initial conditions, as its predictions show considerable divergence from the ground truth and the predicted flow patterns are completely different from the expected ones even at very early rollout stage. By $T=1600$, the model begins to exhibit dispersed small-scale vortices and blurred large-scale structures, with these issues exacerbating over time. This highlights the limitations of purely data-driven model in handling unseen initial conditions despite its strong performance on training conditions. NeuralPDE-Discretize initially displays potential in simulating physical vorticity patterns but struggles with numerical instabilities beyond $T=2400$. The failure of NeuralPDE-Discretize under unseen initial conditions arises from its use of a globally trained CNN to interpolate convection fluxes across the computational domain, which does not generalized well to novel flow patterns. This model integrates trainable neural networks deeply within the numerical solver, specifically in calculating convection flux at cell faces, a process critical for capturing fluid transport and mixing. When exposed to new conditions, the model’s interpolation inaccurately predicts these fluxes due to its reliance on training-specific scenarios. In a FVM setting, where accurate flux calculation is crucial for stability, such errors in flux prediction lead to numerical instability, particularly in convection-dominated chaotic flows where slight inaccuracies can escalate into significant deviations from expected physical behavior. 
\begin{figure}[hpt!]
    \centering
    \includegraphics[width=0.7\textwidth]{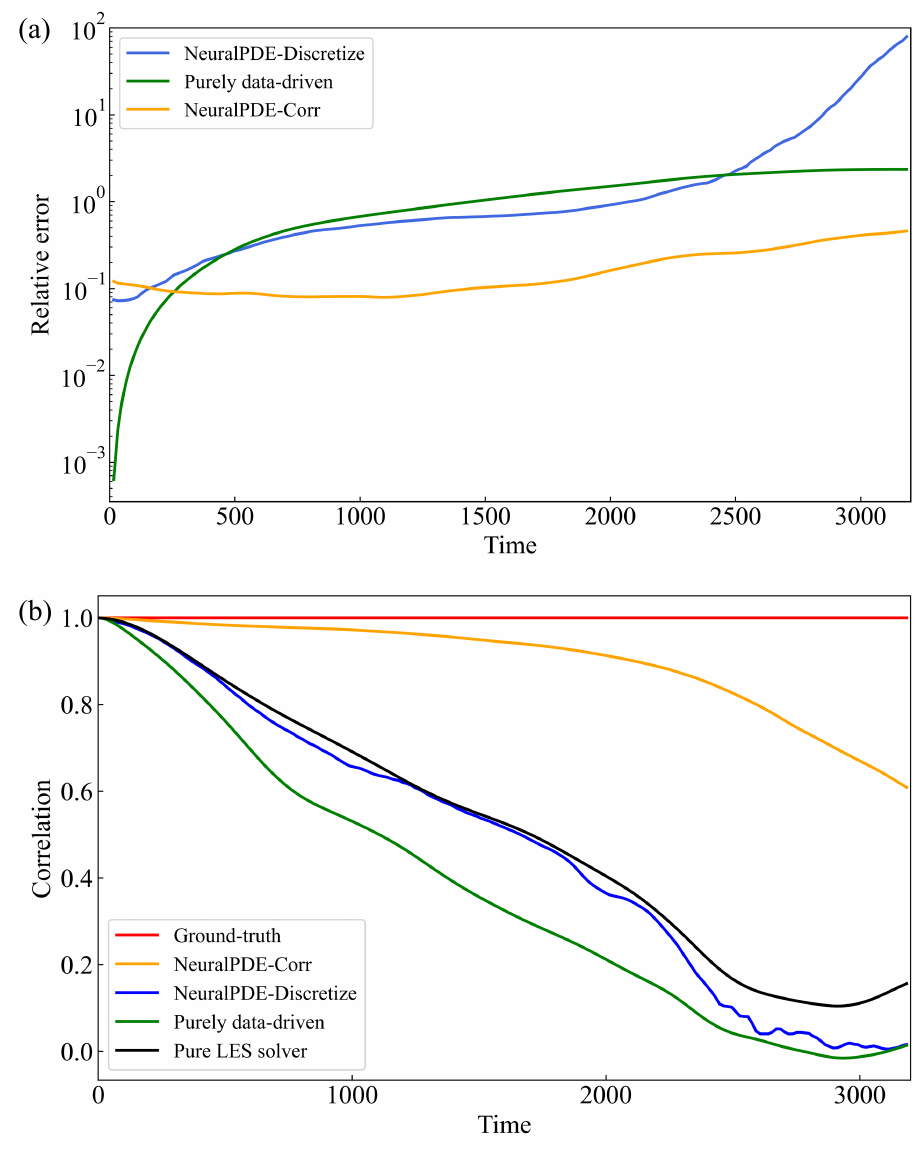}
    \caption{Comparison of (a) relative prediction error and (b) correlation of predicted vorticity with ground truth across all models for unseen initial conditions.}
    \label{fig:error_test}
\end{figure}
The purely physics-based model (i.e., LES) on such coarse space-time resolution demonstrates the most drastic limitations, completely failing to capture the correct dynamics, particularly for the pressure. This is primarily due to insufficient grid resolution and very large timestep. 

Fig.~\ref{fig:error_test} further quantifies the performance of all models under unseen initial conditions through two key metrics: relative prediction error and vorticity correlation with the ground truth. Panel (a) displays the relative errors over time, where NeuralPDE-Corr consistently shows the lowest errors, underlining its superior predictive accuracy. Panel (b) depicts the correlation of predicted vorticity with ground truth, with NeuralPDE-Corr maintaining the highest correlation, suggesting that its predictions are most closely aligned with the true dynamics of turbulence. Despite initial promise, the purely data-driven model shows a marked increase in error and decline in correlation over time, reflecting its inability to generalize beyond the training conditions. Similarly, NeuralPDE-Discretize starts well but soon faces performance degradation due to its numerical stability issues.
\begin{figure}[htp!]
    \centering
    \includegraphics[width=0.65\textwidth]{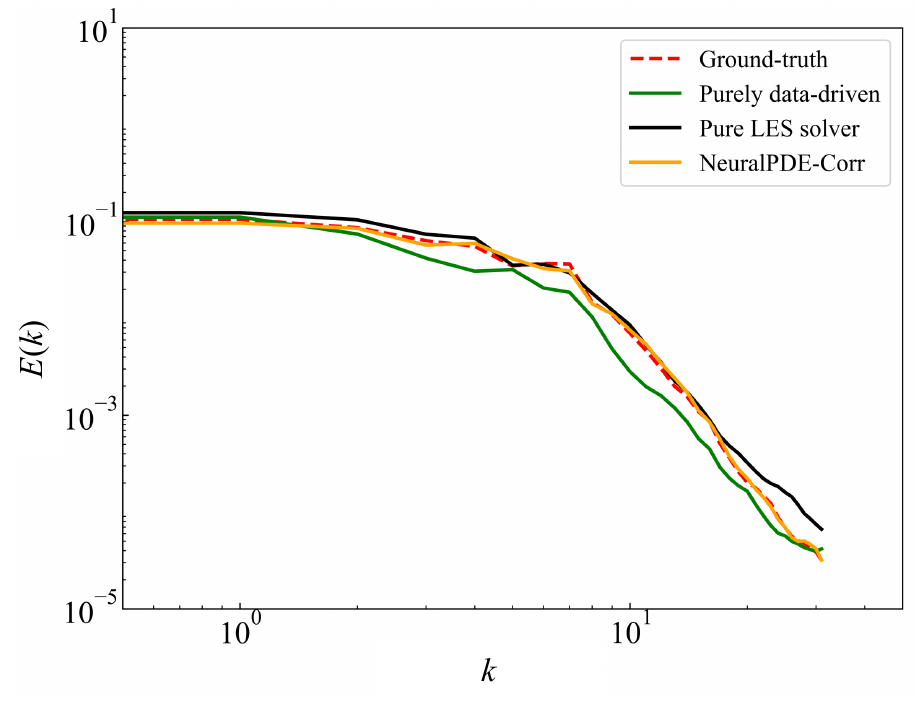}
    \caption{Energy spectra comparison across different models, averaged over the period from $T=2400$ to $T=3200$ to ensure decorrelation from initial conditions. The spectrum for NeuralPDE-Discretize is excluded due to its stability issues.}
    \label{fig:vorticity corr}
\end{figure}
Lastly, the energy spectra comparison is presented in Fig.~\ref{fig:vorticity corr}, which shows how closely each model approaches the true energy distribution across different wave numbers, particularly capturing the energy decay at higher wave numbers. The results further confirm previous observations. 

The comparative analysis of models under unseen initial conditions underscores the promise of seamlessly integrating numerical PDEs into neural networks, exemplified by the hybrid NeuralPDE-Corr model. This hybrid model exhibits superior generalizability and stability over the purely data-driven or conventional numerical solvers by leveraging the strengths of both ML and physics-based approaches. However, the specific design of the hybrid architecture plays a crucial role in determining overall performance, highlighting the need for careful consideration in the configuration of such models to optimize their effectiveness. 

\subsection{Recovered high wavenumber feature by diffusion-SR model}
\label{sec:SR}

\begin{figure}[t!]
    \centering
    \includegraphics[width=\textwidth]{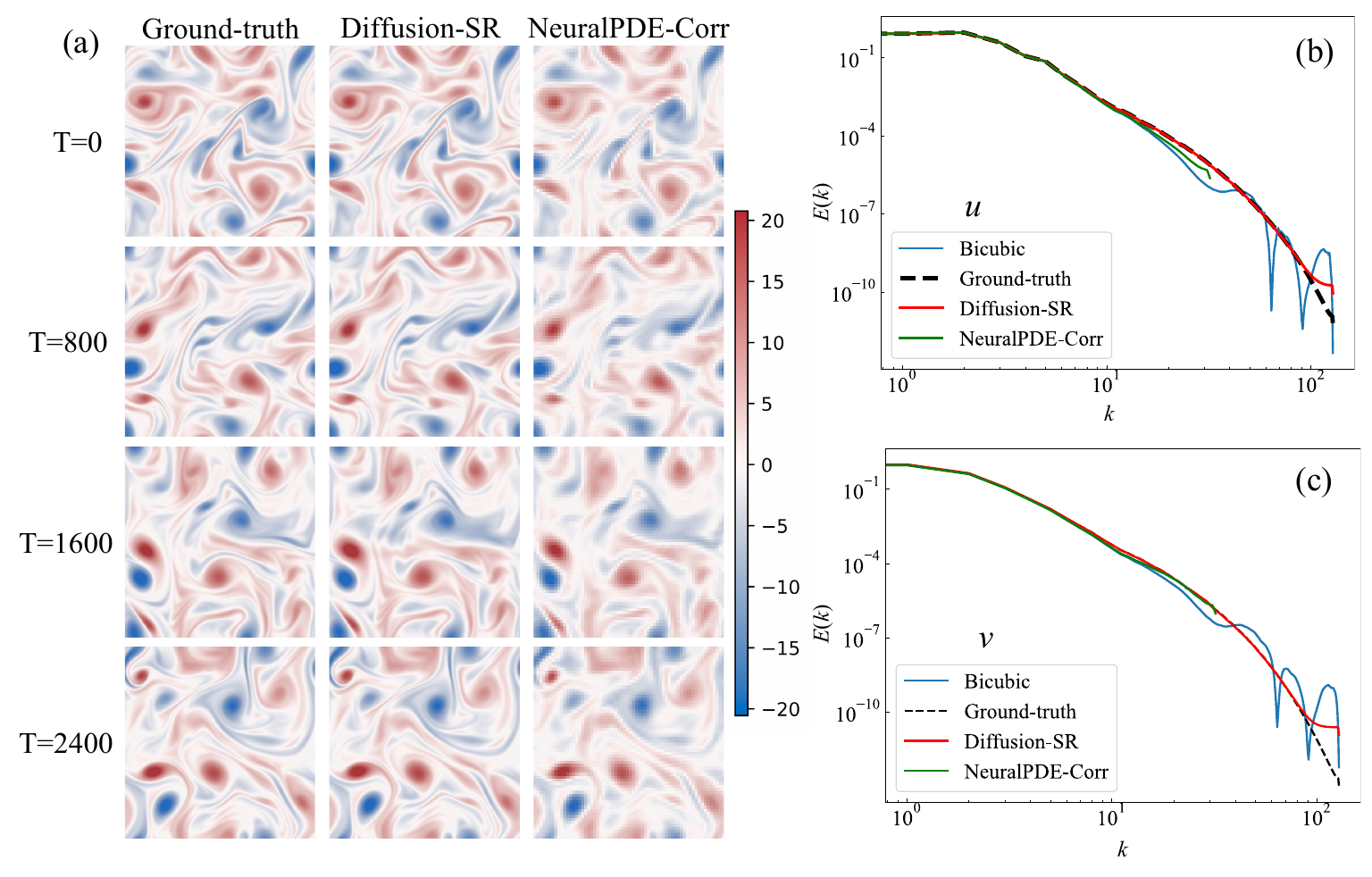}
    \caption{Super resolution by diffusion-SR network: (a) contours of vorticity, where the resolution of PDE-Neural-Corr is $64 \times 64$ and the others are $256 \times 256$; (b) energy spectrum of $\mathbf{u}_x$; (c) energy spectrum of $\mathbf{u}_y$.} \label{fig:SR-vort}
\end{figure}

Conditioned on the low-resolution velocity and pressure data obtained from the hybrid neural solver (NeuralPDE-Corr, $64 \times 64$), the diffusion-SR network outlined in Sec.~\ref{sec:SR-NET} successfully restores high-wavenumber turbulence features and enhances the resolution of flow predictions to $256 \times 256$. Although constrained by current GPU memory to a super-resolution factor of $4\times$, the network has the theoretical capability to handle any higher resolutions if sufficient computational resources are available. The super-resolved velocity and pressure predictions are illustrated in Figs.~\ref{fig:SR-velocity} and ~\ref{fig:SR-p}. We further calculate the vorticity and energy spectrum from the super-resolved velocity fields, as depicted in Fig.~\ref{fig:SR-vort}, to show the SR performance. Overall, the diffusion-SR method not only recaptures but effectively accentuates small-scale turbulence details at higher wavenumbers. The energy spectra in Figs.~\ref{fig:SR-vort} (b) and (c) show that the energy levels ranging from $10^{-10}$ to $10^{-6}$ have been successfully restored, closely matching the energy spectra of the original high-resolution data. This is in contrast to the bicubic interpolation method, which, although simpler, introduces spurious artifacts not present in the actual turbulence data.

In sum, we have showcased the efficacy of the concept of hybrid neural differentiable modeling that integrates physics in the form of discretized governing PDEs with the SOTA machine learning techniques via differentiable programming. The proposed modeling strategy can adeptly capture both large and small scales of turbulent dynamics, while preserving a high level of generalizability across various initial conditions.


\section{Discussions}
\label{sec:discussion}

\subsection{Assessment of enrichment through PDEs}
\label{sec:enrich}

\begin{figure}[htp!]
    \centering
    \includegraphics[width=0.7\textwidth]{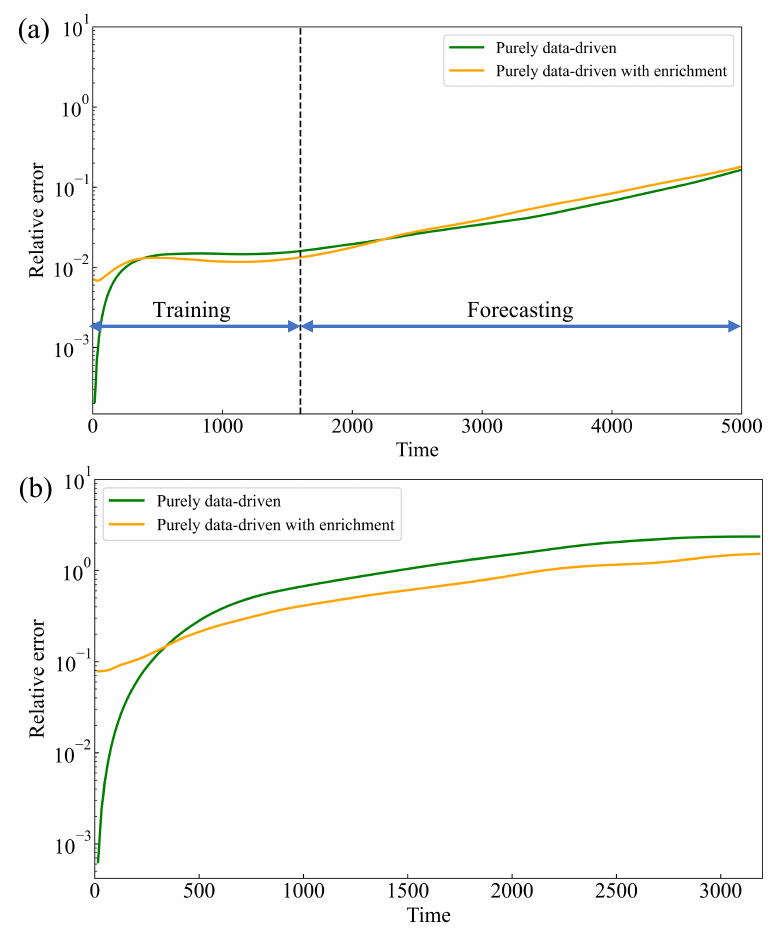}
    \caption{Relative error comparison of purely data-driven model w and w/o PDE feature enrichment, tested on (a) training initial conditions (b) unseen initial conditions.}
    \label{fig:enrichment}
\end{figure}

In hybrid neural models, a important portion of the architecture dedicate itself to non-trainable, discretized physics, enhancing the model's generalizability compared to purely trainable neural networks. These hybrid differentiable models benefit from end-to-end training via differentiable programming, which ensure both accuracy and stability in $a ~ posterior$ evaluations. Disjointed physics and neural networks often encounter significant stability challenges across extended spatiotemporal dynamics, as evidenced in the literature~\cite{fan2024differentiable, um2020solver, wu2019reynolds}. To further illustrate the effectiveness of hybrid neural models from different perspectives, an evaluation is conducted to assess the performance of purely data-driven models when they are augmented with physical features derived from the solver by input feature enrichment. This assessment will provide a comprehensive understanding of the advantages offered by hybrid neural models. For purely data-driven model, if the physics are only used to enrich the input features, the autogressive sequential net can be expressed as,
\begin{equation}
\bm{V}_t = \bm{V}_{t-1} +  \mathrm{ConvLSTM}\Big(\bm{V}_{t-1}, \mathrm{ConvPDE}(\bm{V}_{t-1}), \bm{H}_{t-1}; \bm{\theta}_{nn}\Big),
\end{equation}
The neural architecture and hyperparameters is kept the identical to those of purely data-driven models. The relative errors for both training and unseen ICs are presented in Fig.~\ref{fig:enrichment}. 

In the context of training initial conditions, the relative error trends of both purely data-driven models—with and without physics enrichment—are nearly identical, effectively capturing the spatiotemporal dynamics of the true data. However, when subjected to unseen initial conditions, the model enriched with physics-derived features demonstrates a slight reduction in error, highlighting its enhanced generalizability, although the improvement is modest. In comparison with the hybrid neural models, particularly the NeuralPDE-Corr variant, the error remains an order of magnitude higher even with physics enrichment. The physics incorporated serves to augment the input features of the purely data-driven model, but the overall structure remains predominantly black-box in nature. This underscores the superior performance and generalizability of the proposed hybrid neural models, which seamlessly integrate discretized physics with state-of-the-art ML techniques through differentiable programming, offering substantial advancements in both predictive accuracy and generalizability.



\subsection{Offline training and online online inference cost analysis}
\label{sec:cost}
We conducted a comparative evaluation of the computational cost of different models.  Table~\ref{tab:cost} summarize the cost for simulating 2D Kolmogorov turbulence over an eight-second duration during the inference phase. Notably, our trained hybrid neural model achieved a substantial speedup, operating 26 times faster than DNS, which is itself an optimized GPU-accelerated, fully vectorized CFD solver in \cite{kochkov2021machine}. This speed advantage would likely increase further when compared to established CPU-based CFD solvers like OpenFOAM, as reported in applications of fluid-structure interaction applications~\cite{fan2024differentiable}.
\begin{table}[htp!]
    \centering
    \begin{tabular}{c c c}
    \toprule[1.5pt] 
    DNS & Hybrid neural solver & Purely data-driven model \\ \midrule
      52.74   &  2.03 & 2.31\\
  \bottomrule[1.5pt] 
    \end{tabular}
    \caption{The inference cost of hybrid neural solver, compared with DNS and purely data-driven model. Note that all inference is performed on an RTX A6000 GPU.}
    \label{tab:cost}
\end{table}

The training cost for hybrid models, while comparable to running a high-fidelity DNS simulation on OpenFOAM using 8 CPUs, is a one-time offline investment. Subsequent uses of the trained model for predictions incur minimal costs, making this approach highly advantageous for applications requiring extended temporal forecasting and repetitive analysis across various conditions, such as different Reynolds numbers and initial states. It should be noted that the training cost associated with the hybrid neural model is higher than that of the purely data-driven model, as reported in our previous work \cite{fan2024differentiable}. This additional training overhead primarily arises from two factors. Firstly, the hybrid model demands more memory and training time for back-propagation due to the necessity of tracing gradients through complex, differentiable physics modules. Secondly, to address numerical instabilities inherent to the training process, we may implement transfer learning strategies that progressively increase the length of autoregressive steps during end-to-end training. While these methods elevate the offline training costs, they are justified by the significant improvements in inference performance and generalizability demonstrated in Section \ref{sec: testing}.

\section{Conclusions}
\label{sec:conclusion}
In this paper, we introduced an innovative neural differentiable modeling framework aimed at enhancing turbulence simulation capabilities. Within this concept, we developed a hybrid differentiable neural solver configured on a coarse grid, capturing large-scale turbulent phenomena effectively. Subsequently, we applied a Bayesian conditional diffusion model to generate small-scale turbulence details, conditioned on the large-scale flow predictions. We engineered two unique hybrid architectural prototypes and evaluated their performance in simulating 2D Kolmogorov turbulence, comparing them against traditional purely data-driven models and physics-based models with conventional SGS closures.

Our comparative analysis revealed that the proposed hybrid neural models, comprehensively trained via differentiable programming, outperform all baseline models in accuracy and generalizability. The first architectural design, which appends additional trainable neural networks to the numerical PDE layers, and the second, which deeply merges neural networks with PDEs to learn flux interpolation schemes, both demonstrated the ability to predict turbulence accurately over extended periods, demonstrating their superiority over the baseline models. However, the second design displayed significant numerical instability when applied to unseen initial conditions, highlighting challenges in its generalizability. To mitigate these issues, further constraints and scaling techniques are needed. Furthermore, the proposed diffusion-based super-resolution network successfully reconstructs high-wavenumber turbulence features from low-resolution outputs generated by the coarse hybrid neural solver, effectively demonstrating the framework's capacity to simulate high-fidelity turbulence across all scales. Additionally, various aspects of the model construction and training have been analyzed and discussed,  providing valuable insights into optimizing training efficiency and predictive accuracy. Our findings underscore the potential of hybrid neural differentiable modeling in simulating turbulence, showcasing their superior performance and architectural flexibility.

Looking forward, the insights gained from this study pave the way for further exploration into seamlessly integrating cutting-edge ML with classic CFD techniques via differentiable programming, aiming to advance the capabilities of next-generation data-augmented neural solvers. Future research will focus on expanding these methodologies to more complex applications, illustrating the robustness and adaptability of the neural differentiable framework in tackling real-world engineering challenges.

\section*{Acknowledgements}
 The authors would like to acknowledge the funds from Office of Naval Research under award numbers N00014-23-1-2071 and National Science Foundation under award numbers OAC-2047127. XF would also like to acknowledge the fellowship provided by the Environmental Change Initiative and Center for Sustainable Energy at University of Notre Dame. 
 
\section*{Compliance with Ethical Standards}
Conflict of Interest: The authors declare that they have no conflict of interest.

\appendix
\section{Details of each hybrid neural models}

\begin{figure}[H]
    \centering
    \includegraphics[width=\textwidth]{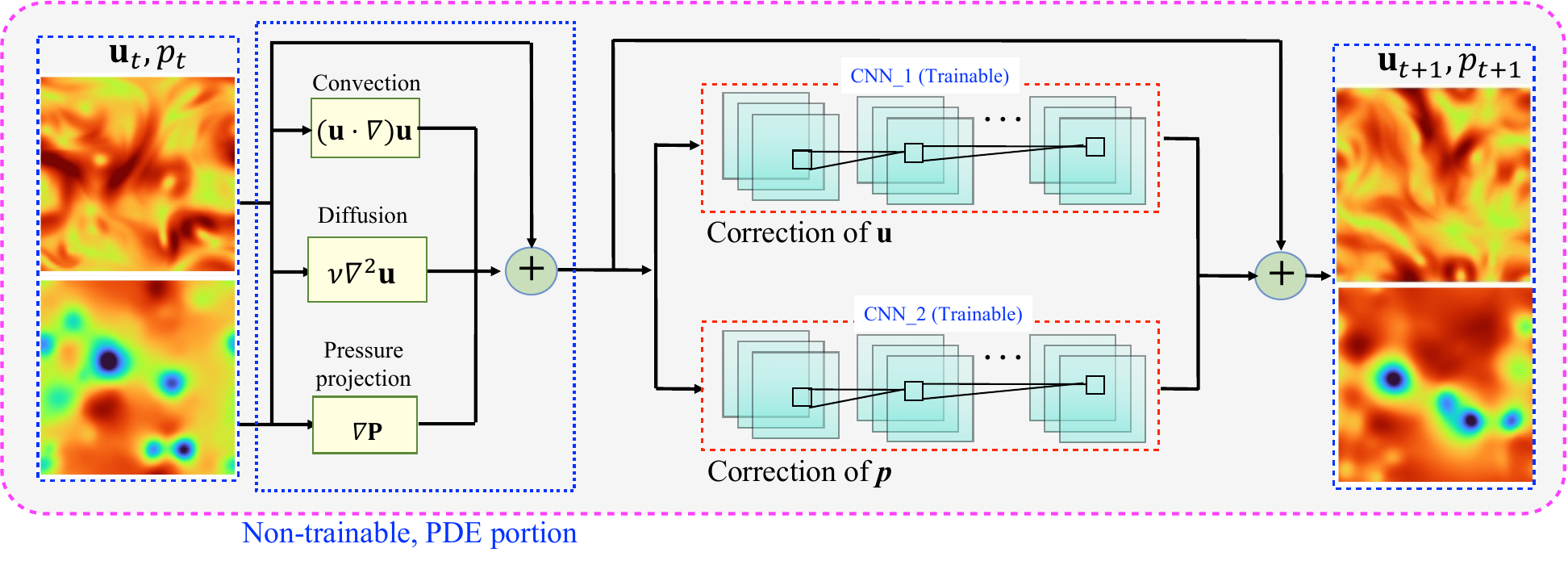}
    \caption{Details of NeuralPDE-Corr in one learning step.}
    \label{fig:details_correct}
\end{figure}

\begin{figure}[H]
    \centering
    \includegraphics[width=\textwidth]{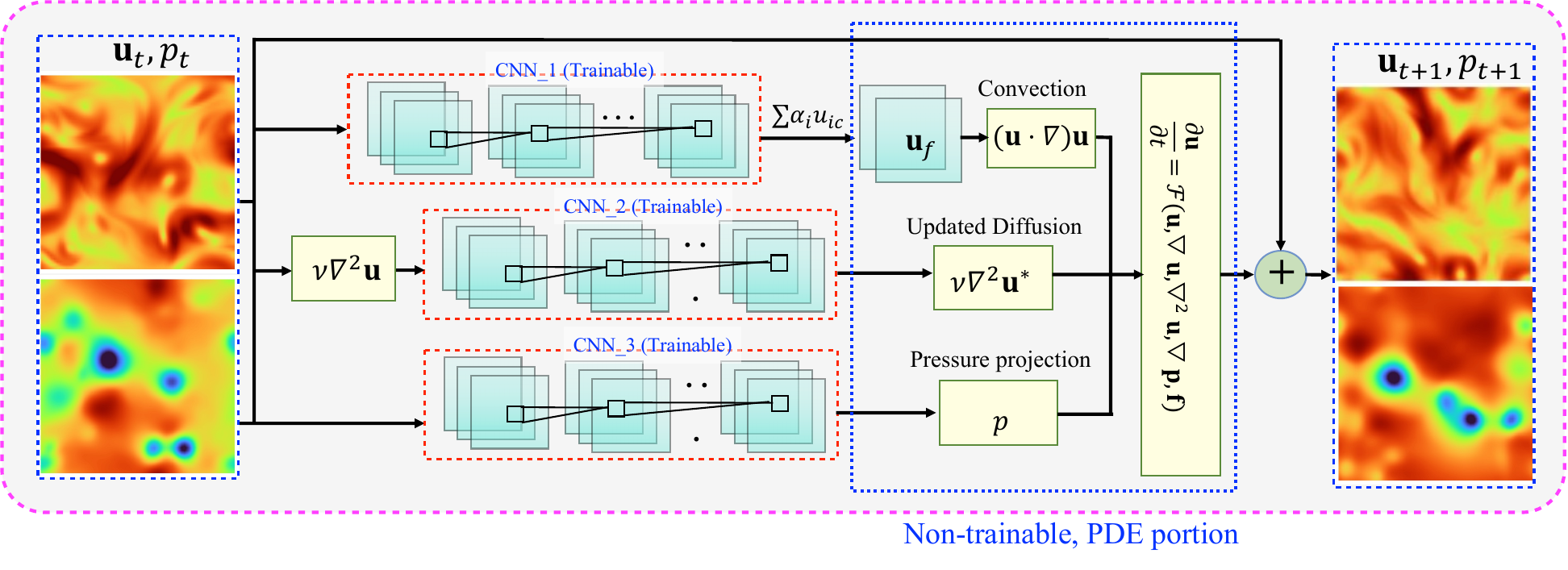}
    \caption{Details of NeuralPDE-Discretize in one learning step.}
    \label{fig:details_num}
\end{figure}


\section{Super resolution of velocity and pressure}

\begin{figure}[H]
    \centering
    \includegraphics[width=0.9\textwidth]{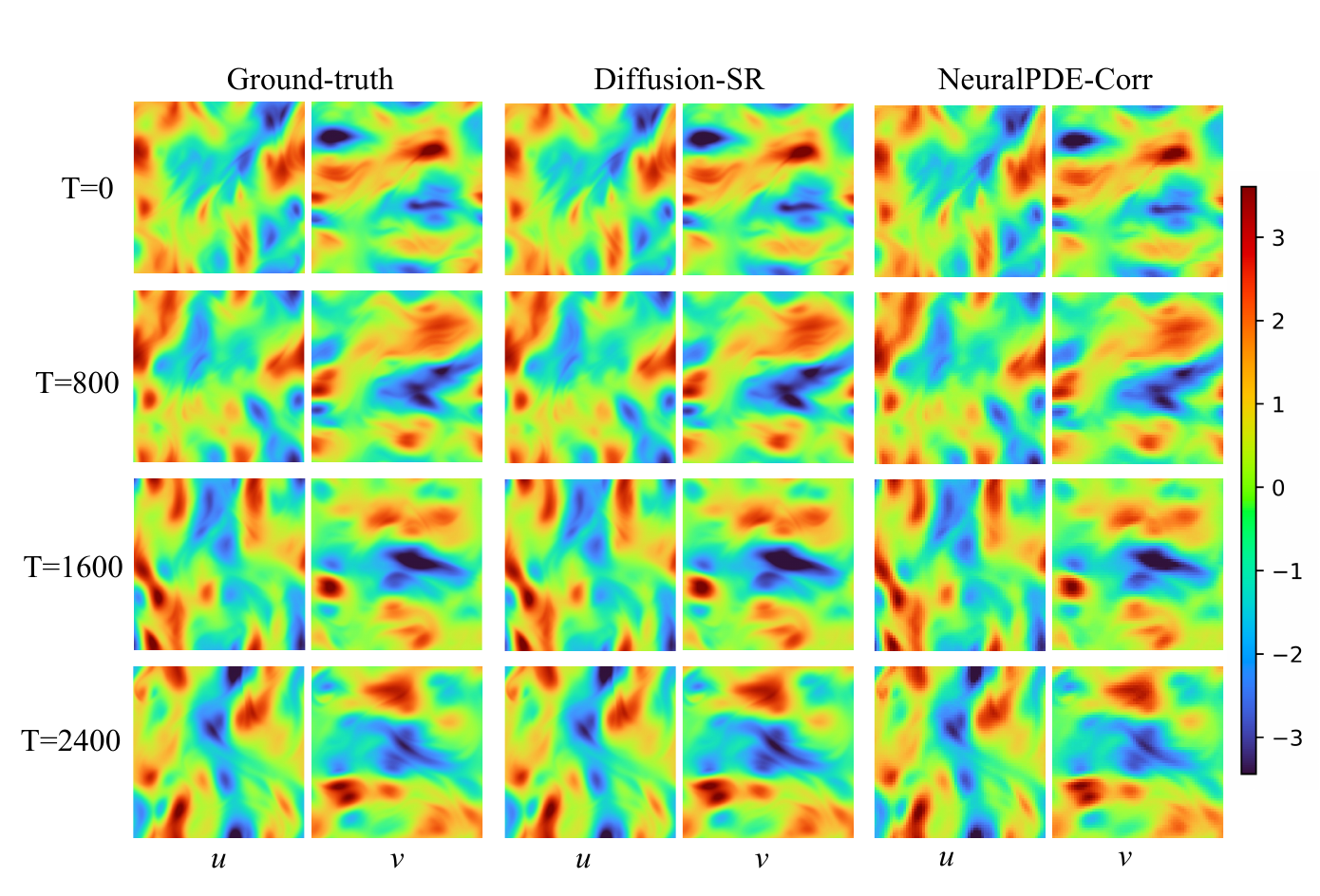}
    \vspace{-1em}
    \caption{Super resolution of velocity by diffusion-SR network.} 
    \label{fig:SR-velocity}
\end{figure}

\begin{figure}[H]
    \centering
    \includegraphics[width=0.9\textwidth]{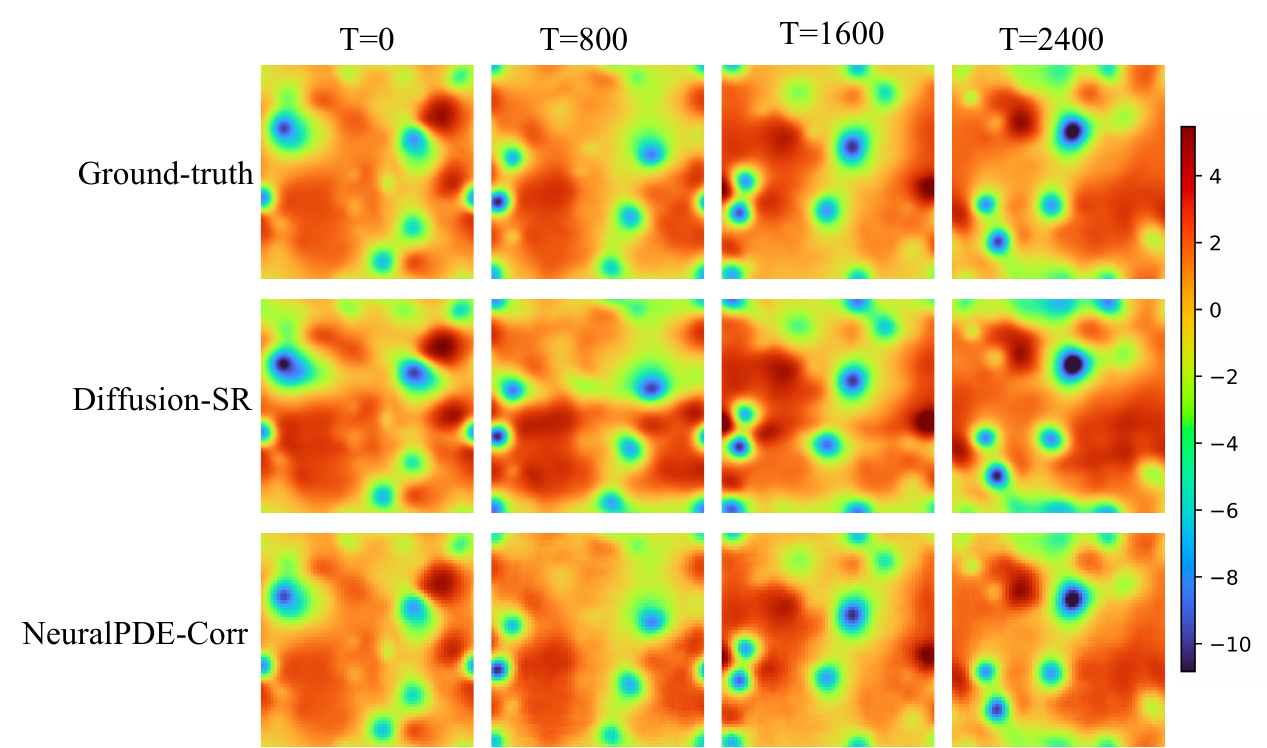}
    \caption{Super resolution of pressure by diffusion-SR network.} 
    \label{fig:SR-p}
\end{figure}

\bibliographystyle{elsarticle-num}
\bibliography{ref,own-ref}

\begin{thebibliography}{10}
\expandafter\ifx\csname url\endcsname\relax
  \def\url#1{\texttt{#1}}\fi
\expandafter\ifx\csname urlprefix\endcsname\relax\def\urlprefix{URL }\fi
\expandafter\ifx\csname href\endcsname\relax
  \def\href#1#2{#2} \def\path#1{#1}\fi

\bibitem{durbin2018some}
P.~A. Durbin, Some recent developments in turbulence closure modeling, Annual
  Review of Fluid Mechanics 50 (2018) 77--103.

\bibitem{morimoto2021convolutional}
M.~Morimoto, K.~Fukami, K.~Zhang, A.~G. Nair, K.~Fukagata, Convolutional neural
  networks for fluid flow analysis: toward effective metamodeling and low
  dimensionalization, Theoretical and Computational Fluid Dynamics 35~(5)
  (2021) 633--658.

\bibitem{guastoni2021convolutional}
L.~Guastoni, A.~G{\"u}emes, A.~Ianiro, S.~Discetti, P.~Schlatter, H.~Azizpour,
  R.~Vinuesa, Convolutional-network models to predict wall-bounded turbulence
  from wall quantities, Journal of Fluid Mechanics 928 (2021) A27.

\bibitem{pfaff2020learning}
T.~Pfaff, M.~Fortunato, A.~Sanchez-Gonzalez, P.~W. Battaglia, Learning
  mesh-based simulation with graph networks, in: International Conference on
  Learning Representations, 2020.

\bibitem{han2022predicting}
X.~Han, H.~Gao, T.~Pfaff, J.-X. Wang, L.~Liu, Predicting physics in
  mesh-reduced space with temporal attention, in: International Conference on
  Learning Representations, 2022.

\bibitem{li2020fourier}
Z.~Li, N.~Kovachki, K.~Azizzadenesheli, B.~Liu, K.~Bhattacharya, A.~Stuart,
  A.~Anandkumar, Fourier neural operator for parametric partial differential
  equations, arXiv preprint arXiv:2010.08895 (2020).

\bibitem{li2023long}
Z.~Li, W.~Peng, Z.~Yuan, J.~Wang, Long-term predictions of turbulence by
  implicit u-net enhanced fourier neural operator, Physics of Fluids 35~(7)
  (2023).

\bibitem{lu2021learning}
L.~Lu, P.~Jin, G.~Pang, Z.~Zhang, G.~E. Karniadakis, Learning nonlinear
  operators via deeponet based on the universal approximation theorem of
  operators, Nature machine intelligence 3~(3) (2021) 218--229.

\bibitem{demo2023deeponet}
N.~Demo, M.~Tezzele, G.~Rozza, A deeponet multi-fidelity approach for residual
  learning in reduced order modeling, arXiv preprint arXiv:2302.12682 (2023).

\bibitem{karniadakis2021physics}
G.~E. Karniadakis, I.~G. Kevrekidis, L.~Lu, P.~Perdikaris, S.~Wang, L.~Yang,
  Physics-informed machine learning, Nature Reviews Physics 3~(6) (2021)
  422--440.

\bibitem{cai2021physics}
S.~Cai, Z.~Mao, Z.~Wang, M.~Yin, G.~E. Karniadakis, Physics-informed neural
  networks (pinns) for fluid mechanics: A review, Acta Mechanica Sinica 37~(12)
  (2021) 1727--1738.

\bibitem{raissi2019physics}
M.~Raissi, P.~Perdikaris, G.~E. Karniadakis, Physics-informed neural networks:
  A deep learning framework for solving forward and inverse problems involving
  nonlinear partial differential equations, Journal of Computational physics
  378 (2019) 686--707.

\bibitem{sun2020surrogate}
L.~Sun, H.~Gao, S.~Pan, J.-X. Wang, Surrogate modeling for fluid flows based on
  physics-constrained deep learning without simulation data, Computer Methods
  in Applied Mechanics and Engineering 361 (2020) 112732.

\bibitem{sun2022bayesian}
L.~Sun, D.~Z. Huang, H.~Sun, J.-X. Wang, Bayesian spline learning for equation
  discovery of nonlinear dynamics with quantified uncertainty, in: NeurIPS,
  PMLR, 2022.

\bibitem{sun2020physics}
L.~Sun, J.-X. Wang, Physics-constrained bayesian neural network for fluid flow
  reconstruction with sparse and noisy data, Theoretical and Applied Mechanics
  Letters 10~(3) (2020) 161--169.

\bibitem{gao2021phygeonet}
H.~Gao, L.~Sun, J.-X. Wang, {PhyGeoNet:} physics-informed geometry-adaptive
  convolutional neural networks for solving parameterized steady-state {PDEs}
  on irregular domain, Journal of Computational Physics 428 (2021) 110079.

\bibitem{gao2022physics}
H.~Gao, M.~J. Zahr, J.-X. Wang, Physics-informed graph neural galerkin
  networks: A unified framework for solving pde-governed forward and inverse
  problems, Computer Methods in Applied Mechanics and Engineering 390 (2022)
  114502.

\bibitem{ren2021phycrnet}
P.~Ren, C.~Rao, Y.~Liu, J.-X. Wang, H.~Sun, Phycrnet: Physics-informed
  convolutional-recurrent network for solving spatiotemporal pdes, Computer
  Methods in Applied Mechanics and Engineering 389 (2022) 114399.

\bibitem{ren2024seismicnet}
P.~Ren, C.~Rao, S.~Chen, J.-X. Wang, H.~Sun, Y.~Liu, Seismicnet:
  Physics-informed neural networks for seismic wave modeling in semi-infinite
  domain, Computer Physics Communications 295 (2024) 109010.

\bibitem{arzani2021uncovering}
A.~Arzani, J.-X. Wang, R.~M. D'Souza, Uncovering near-wall blood flow from
  sparse data with physics-informed neural networks, Physics of Fluids 33~(7)
  (2021) 071905.
\newblock \href {https://doi.org/10.1063/5.0055600}
  {\path{doi:10.1063/5.0055600}}.

\bibitem{movahhedi2023predicting}
M.~Movahhedi, X.-Y. Liu, B.~Geng, C.~Elemans, Q.~Xue, J.-X. Wang, X.~Zheng,
  Predicting 3d soft tissue dynamics from 2d imaging using physics informed
  neural networks, Communications Biology 6~(1) (2023) 541.

\bibitem{li2022physics}
R.~Li, J.-X. Wang, E.~Lee, T.~Luo, Physics-informed deep learning for solving
  phonon boltzmann transport equation with large temperature non-equilibrium,
  npj computational materials 8 (2022) 19.

\bibitem{kharazmi2021inferring}
E.~Kharazmi, D.~Fan, Z.~Wang, M.~S. Triantafyllou, Inferring vortex induced
  vibrations of flexible cylinders using physics-informed neural networks,
  Journal of Fluids and Structures 107 (2021) 103367.

\bibitem{krishnapriyan2021characterizing}
A.~Krishnapriyan, A.~Gholami, S.~Zhe, R.~Kirby, M.~W. Mahoney, Characterizing
  possible failure modes in physics-informed neural networks, Advances in
  Neural Information Processing Systems 34 (2021) 26548--26560.

\bibitem{wang2022and}
S.~Wang, X.~Yu, P.~Perdikaris, When and why pinns fail to train: A neural
  tangent kernel perspective, Journal of Computational Physics 449 (2022)
  110768.

\bibitem{ling2016reynolds}
J.~Ling, A.~Kurzawski, J.~Templeton, Reynolds averaged turbulence modelling
  using deep neural networks with embedded invariance, Journal of Fluid
  Mechanics 807 (2016) 155--166.

\bibitem{weatheritt2017hybrid}
J.~Weatheritt, R.~D. Sandberg, Hybrid reynolds-averaged/large-eddy simulation
  methodology from symbolic regression: formulation and application, AIAA
  Journal 55~(11) (2017) 3734--3746.

\bibitem{wang2017physics}
J.-X. Wang, J.-L. Wu, H.~Xiao, Physics-informed machine learning approach for
  reconstructing reynolds stress modeling discrepancies based on {DNS} data,
  Physical Review Fluids 2~(3) (2017) 034603.

\bibitem{wang2019prediction}
J.-X. Wang, J.~Huang, L.~Duan, H.~Xiao, Prediction of reynolds stresses in
  high-mach-number turbulent boundary layers using physics-informed machine
  learning, Theoretical and Computational Fluid Dynamics 33~(1) (2019) 1--19.

\bibitem{maulik2019subgrid}
R.~Maulik, O.~San, A.~Rasheed, P.~Vedula, Subgrid modelling for two-dimensional
  turbulence using neural networks, Journal of Fluid Mechanics 858 (2019)
  122--144.

\bibitem{yang2019predictive}
X.~Yang, S.~Zafar, J.-X. Wang, H.~Xiao, Predictive large-eddy-simulation wall
  modeling via physics-informed neural networks, Physical Review Fluids 4~(3)
  (2019) 034602.

\bibitem{zhou2021wall}
Z.~Zhou, G.~He, X.~Yang, Wall model based on neural networks for les of
  turbulent flows over periodic hills, Physical Review Fluids 6~(5) (2021)
  054610.

\bibitem{lozano2020self}
A.~Lozano-Dur{\'a}n, H.~J. Bae, Self-critical machine-learning wall-modeled les
  for external aerodynamics, arXiv preprint arXiv:2012.10005 (2020).

\bibitem{wu2019reynolds}
J.~Wu, H.~Xiao, R.~Sun, Q.~Wang, Reynolds-averaged navier--stokes equations
  with explicit data-driven reynolds stress closure can be ill-conditioned,
  Journal of Fluid Mechanics 869 (2019) 553--586.

\bibitem{guan2022stable}
Y.~Guan, A.~Chattopadhyay, A.~Subel, P.~Hassanzadeh, Stable a posteriori les of
  2d turbulence using convolutional neural networks: Backscattering analysis
  and generalization to higher re via transfer learning, Journal of
  Computational Physics 458 (2022) 111090.

\bibitem{mcconkey2021curated}
R.~McConkey, E.~Yee, F.-S. Lien, A curated dataset for data-driven turbulence
  modelling, Scientific data 8~(1) (2021) 255.

\bibitem{zhang2022ensemble}
X.-L. Zhang, H.~Xiao, X.~Luo, G.~He, Ensemble kalman method for learning
  turbulence models from indirect observation data, Journal of Fluid Mechanics
  949 (2022) A26.

\bibitem{baydin2018automatic}
A.~G. Baydin, B.~A. Pearlmutter, A.~A. Radul, J.~M. Siskind, Automatic
  differentiation in machine learning: a survey, Journal of Marchine Learning
  Research 18 (2018) 1--43.

\bibitem{mensch2018differentiable}
A.~Mensch, M.~Blondel, Differentiable dynamic programming for structured
  prediction and attention, in: International Conference on Machine Learning,
  PMLR, 2018, pp. 3462--3471.

\bibitem{innes2019differentiable}
M.~Innes, A.~Edelman, K.~Fischer, C.~Rackauckas, E.~Saba, V.~B. Shah,
  W.~Tebbutt, A differentiable programming system to bridge machine learning
  and scientific computing, arXiv preprint arXiv:1907.07587 (2019).

\bibitem{belbute2020combining}
F.~D.~A. Belbute-Peres, T.~Economon, Z.~Kolter, Combining differentiable pde
  solvers and graph neural networks for fluid flow prediction, in:
  international conference on machine learning, PMLR, 2020, pp. 2402--2411.

\bibitem{kochkov2021machine}
D.~Kochkov, J.~A. Smith, A.~Alieva, Q.~Wang, M.~P. Brenner, S.~Hoyer, Machine
  learning--accelerated computational fluid dynamics, Proceedings of the
  National Academy of Sciences 118~(21) (2021) e2101784118.

\bibitem{list2022learned}
B.~List, L.-W. Chen, N.~Thuerey, Learned turbulence modelling with
  differentiable fluid solvers: physics-based loss functions and optimisation
  horizons, Journal of Fluid Mechanics 949 (2022) A25.

\bibitem{akhare2023physics}
D.~Akhare, T.~Luo, J.-X. Wang, Physics-integrated neural differentiable
  {(PiNDiff)} model for composites manufacturing, Computer Methods in Applied
  Mechanics and Engineering 406 (2023) 115902.

\bibitem{akhare2023diffhybrid}
D.~Akhare, T.~Luo, J.-X. Wang, Diffhybrid-uq: Uncertainty quantification for
  differentiable hybrid neural modeling, arXiv preprint arXiv:2401.00161
  (2023).

\bibitem{akhare2023probabilistic}
D.~Akhare, Z.~Chen, R.~Gulotty, T.~Luo, J.-X. Wang, Probabilistic
  physics-integrated neural differentiable modeling for isothermal chemical
  vapor infiltration process, arXiv preprint arXiv:2311.07798 (2023).

\bibitem{fan2024differentiable}
X.~Fan, J.-X. Wang, Differentiable hybrid neural modeling for fluid-structure
  interaction, Journal of Computational Physics 496 (2024) 112584.

\bibitem{liu2024multi}
X.-Y. Liu, M.~Zhu, L.~Lu, H.~Sun, J.-X. Wang, Multi-resolution partial
  differential equations preserved learning framework for spatiotemporal
  dynamics, Communications Physics 7~(1) (2024) 31.

\bibitem{macart2021embedded}
J.~F. MacArt, J.~Sirignano, J.~B. Freund, Embedded training of neural-network
  subgrid-scale turbulence models, Physical Review Fluids 6~(5) (2021) 050502.

\bibitem{strofer2021end}
C.~A.~M. Str{\"o}fer, H.~Xiao, End-to-end differentiable learning of turbulence
  models from indirect observations, Theoretical and Applied Mechanics Letters
  11~(4) (2021) 100280.

\bibitem{shankar2023differentiable}
V.~Shankar, V.~Puri, R.~Balakrishnan, R.~Maulik, V.~Viswanathan, Differentiable
  physics-enabled closure modeling for burgers’ turbulence, Machine Learning:
  Science and Technology (2023).

\bibitem{shankar2023differentiableturbulence}
V.~Shankar, R.~Maulik, V.~Viswanathan, Differentiable turbulence ii, arXiv
  preprint arXiv:2307.13533 (2023).

\bibitem{mccomb2014homogeneous}
W.~D. McComb, Homogeneous, Isotropic Turbulence: Phenomenology, Renormalization
  and Statistical Closures, Vol. 162, OUP Oxford, 2014.

\bibitem{chandler2013invariant}
G.~J. Chandler, R.~R. Kerswell, Invariant recurrent solutions embedded in a
  turbulent two-dimensional kolmogorov flow, Journal of Fluid Mechanics 722
  (2013) 554--595.

\bibitem{bar2019learning}
Y.~Bar-Sinai, S.~Hoyer, J.~Hickey, M.~P. Brenner, Learning data-driven
  discretizations for partial differential equations, Proceedings of the
  National Academy of Sciences 116~(31) (2019) 15344--15349.

\bibitem{zhu2019machine}
L.~Zhu, W.~Zhang, J.~Kou, Y.~Liu, Machine learning methods for turbulence
  modeling in subsonic flows around airfoils, Physics of Fluids 31~(1) (2019)
  015105.

\bibitem{gao2023bayesian}
H.~Gao, X.~Han, X.~Fan, L.~Sun, L.-P. Liu, L.~Duan, J.-X. Wang, Bayesian
  conditional diffusion models for versatile spatiotemporal turbulence
  generation, Computer Methods in Applied Mechanics and Engineering 427 (2024)
  117023.

\bibitem{du2024confild}
P.~Du, M.~H. Parikh, X.~Fan, X.-Y. Liu, J.-X. Wang, {CoNFiLD}: Conditional
  neural field latent diffusion model generating spatiotemporal turbulence,
  arXiv preprint arXiv:2403.05940 (2024).

\bibitem{dhariwal2021diffusion}
P.~Dhariwal, A.~Nichol, Diffusion models beat gans on image synthesis, Advances
  in neural information processing systems 34 (2021) 8780--8794.

\bibitem{ho2020denoising}
J.~Ho, A.~Jain, P.~Abbeel, Denoising diffusion probabilistic models, Advances
  in neural information processing systems 33 (2020) 6840--6851.

\bibitem{chung2022diffusion}
H.~Chung, J.~Kim, M.~T. Mccann, M.~L. Klasky, J.~C. Ye, Diffusion posterior
  sampling for general noisy inverse problems, arXiv preprint arXiv:2209.14687
  (2022).

\bibitem{um2020solver}
K.~Um, R.~Brand, Y.~R. Fei, P.~Holl, N.~Thuerey, Solver-in-the-loop: Learning
  from differentiable physics to interact with iterative pde-solvers, Advances
  in Neural Information Processing Systems 33 (2020) 6111--6122.

\bibitem{gao2021bi}
H.~Gao, J.-X. Wang, A bi-fidelity ensemble kalman method for {PDE-constrained}
  inverse problems in computational mechanics, Computational Mechanics 67~(4)
  (2021) 1115--1131.

\end{thebibliography}

\end{document}